\documentclass[prd,nofootinbib,preprint,superscriptaddress]{revtex4}
\usepackage{amsmath, amssymb, amsthm, graphicx, epsfig, fancyhdr,epsfig, slashed, mathrsfs}
\usepackage{braket}
\usepackage{tikzsymbols}
\usepackage{tikz}
\usepackage{tikz-feynman}
\tikzfeynmanset{compat=1.1.0}
\usepackage{natbib}
\usepackage{float}
\usetikzlibrary{shadows}
\usepackage{comment}
\usepackage{pifont}
\usepackage{booktabs}
\usetikzlibrary{arrows.meta, positioning, shapes, decorations.markings}
\usepackage{caption}
\usepackage{adjustbox}
\usepackage{xcolor}
\usepackage{mathtools}   

\usepackage{amsthm}

\theoremstyle{plain}

\theoremstyle{definition}

\theoremstyle{remark}

\usepackage{physics}   
\usepackage{bm}        
\usepackage{hyperref} 
\usepackage{tikz,xcolor,hyperref}
\usepackage{subfig}

\definecolor{lime}{HTML}{A6CE39}
\DeclareRobustCommand{\orcidicon}{
	\begin{tikzpicture}
	\draw[lime, fill=lime] (0,0) 
	circle [radius=0.2] 
	node[white] {{\fontfamily{qag}\selectfont \tiny ID}};
	\draw[white, fill=white] (-0.0625,0.095) 
	circle [radius=0.007];
	\end{tikzpicture}
	\hspace{-2mm}
}

\foreach \x in {A, ..., Z}{\expandafter\xdef\csname orcid\x\endcsname{\noexpand\href{https://orcid.org/\csname orcidauthor\x\endcsname}
			{\noexpand\orcidicon}}
}

\newcommand{\be}{\begin{equation}}
\newcommand{\ee}{\end{equation}}
\newcommand{\bea}{\begin{eqnarray}}
\newcommand{\eea}{\end{eqnarray}}

\newcommand{\beq}{\begin{equation}}
\newcommand{\eeq}{\end{equation}}

\begin{document}



\title{Cosmic Strings Gravitational Wave Probe of Leptogenesis: \it{Thermal, Non-thermal, Near-resonant and Flavourful}}

\author{Anish Ghoshal\orcidB{}}
\email{anish.ghoshal@fuw.edu.pl}
\affiliation{Department of Physics and Astronomy, University of Sussex, Brighton, BN1 9RH, United Kingdom}

\author{Angus Spalding \orcidC{}}
\email{angus.spalding1@gmail.com}
\affiliation{School of Physics and Astronomy, University of Southampton,
Southampton SO17 1BJ, United Kingdom}

\author{Graham White \orcidD{}}
\email{graham.white@gmail.com}
\affiliation{School of Physics and Astronomy, University of Southampton,
Southampton SO17 1BJ, United Kingdom}

\begin{abstract}
Breaking a global or local $U(1)_{\rm B-L}$ symmetry at high scales simultaneously generates Majorana masses for heavy right-handed neutrinos and produces a network of cosmic strings. The evolution and decay of these strings source a stochastic gravitational-wave background that may be probed by current and future gravitational-wave experiments, while the decays of the resulting massive right-handed neutrinos can generate the baryon asymmetry of the Universe via leptogenesis. We derive analytical bounds for successful leptogenesis with a global and a local $U(1)_{B-L}$ symmetry, separately finding an absolute lower bound on the lightest right-handed neutrino mass $M_1 > 1.74 \times 10^{8}\,\mathrm{GeV}$ for thermal initial conditions and $M_1 > \mathcal{O}(10^{6})\,\mathrm{GeV}$ for non-thermal initial conditions. Allowing for near-resonant leptogenesis relaxes these bounds to TeV scale in both cases making it a viable target at collider searches complementing the GW signals. Full flavour effects are included, and crucially, we determine the region where successful leptogenesis can be probed through gravitational-wave observations in upcoming experiments such as LISA and Einstein Telescope. Importantly, we find that flavour effects rescue regions of the parameter space that are ruled out due to current CMB or gravitational wave measurements.
\end{abstract}

\maketitle

\tableofcontents
\section{Introduction}
The discovery of neutrino oscillations \cite{Fukuda_2001, Fukuda_2002, Ahmad_2002, Ashie_2005, Abe_2016, Smirnov_2016, Aartsen_2018, Abe_2008, Abe_2011} demonstrated that neutrinos possess non-zero masses, in clear contradiction with the Standard Model. Explaining the origin and smallness of these masses therefore calls for new physics beyond the Standard Model. Among the proposed mechanisms, the Type-I seesaw mechanism \cite{Minkowski:1977sc, Gell-Mann:1979vob, Yanagida:1979as, Mohapatra:1980yp} stands out as one of the most elegant and economical explanations. In this framework, the Standard Model is extended by heavy right-handed neutrinos that do not participate in standard model gauge interactions. Through Yukawa interactions with the Higgs field, the new heavy states couple to the left-handed neutrinos; once the Higgs acquires its vacuum expectation value, these interactions generate light neutrino masses inversely proportional to the heavy mass scale. This seesaw relationship naturally accounts for the tiny observed neutrino masses: if the right-handed neutrinos lie at a very high mass scale, the resulting light neutrinos become correspondingly light, even for Yukawa couplings of moderate size. \\
A particularly well-motivated origin for the large Majorana masses of the right-handed neutrinos arises in theories with a $U(1)_{B-L}$ symmetry, corresponding to the difference between baryon and lepton number. When the $U(1)_{B-L}$ symmetry is spontaneously broken at a high scale, the associated scalar field acquires a vacuum expectation value that generates large Majorana masses for the right-handed neutrinos through their Yukawa interactions. The heaviness of these states therefore directly reflects the high scale of $B-L$ breaking, and the use of a symmetry explains the hierarchy between the GUT scale and the scale of neutrinos needed to avoid issues with perturbativity \cite{Gell-Mann:1979vob, Dror:2019syi, Dunsky:2021tih, Ipek:2018sai}. \\
Along with generating the heavy right-handed neutrino masses, transitions \cite{Mazumdar_2019, Athron:2023xlk, quiros1999finitetemperaturefieldtheory} associated with the spontaneous breaking of global $U(1)_{B-L}$ symmetry tend to lead to the formation of cosmic strings \cite{Vilenkin:2000jqa}. These one-dimensional topological defects survive long after their formation and emit gravitational radiation through the production and decay of string loops, sourcing a stochastic gravitational-wave background (GWB) \cite{Fu:2023nrn, King:2020hyd, Ghoshal:2023sfa, Dror:2019syi}.\par

A key feature of cosmic strings is that they are long-standing GW sources \cite{Vilenkin:1981bx,Vachaspati:1984gt,  Hindmarsh:1994re, Vilenkin:2000jqa}. Once the network of cosmic strings is produced, it is expected to quickly enter a scaling regime where they occupy a constant fraction of the total energy density of the universe \cite{Vilenkin:2000jqa, Albrecht:1984xv, Bennett:1987vf, Allen:1990tv,Martins:2000cs, Figueroa:2012kw, Martins:2016ois}. This leads to GW emissions occurring during almost all along the the entire universe history. This generates a near uniform GW spectrum spanning several orders of magnitude in frequencies.\footnote{Specifically, during radiation domination local strings only depart from a uniform distribution due to evolution in $g_\ast$ whereas global strings have a slight logarithmic dependence} A measurement of the GW spectrum from lower to higher frequencies would thus allow to infer about the universe expansion rate from early to later times \cite{Cui:2017ufi, Cui:2018rwi,  Gouttenoire:2019kij,Gouttenoire:2019rtn,Blasi:2020mfx}.\footnote{This for instance is different that typical short-lived GW sources like the ones from first-order phase transitions which may only probe the cosmic history within a small window of time around when the phase transition source typically bubble walls is active \cite{Barenboim:2016mjm,Hook:2020phx,Ellis:2020nnr,Gouttenoire:2021jhk,Domenech:2020kqm}.}\footnote{
 Intermediate cosmological eras can imprint signatures on inflationary GW spectrum as well but the GW amplitude may be too small to be detected in standard inflationary scenarios due to already existing bounds from CMB measurements on the PGW spectrum \cite{DEramo:2019tit, Bernal:2020ywq,Gouttenoire:2021jhk,Gouttenoire:2021wzu,Dunsky:2021tih,Berbig:2023yyy}.} GW backgrounds generated by cosmic strings therefore constitute a unique observational window into the expansion rate of the early Universe and whether there was any departure from standard radiation domination.

The same heavy right-handed neutrinos responsible for generating light neutrino masses in the Type-I seesaw can also play a crucial role in explaining the observed baryon asymmetry of the Universe through leptogenesis \cite{Fukugita:1986hr}. In this framework, the out-of-equilibrium and $CP$-violating decays of the lightest right-handed neutrino generate a net lepton asymmetry in the early Universe. This asymmetry is subsequently reprocessed into a baryon asymmetry by electroweak sphalerons \cite{Luty:1992un, Giudice_2004, Covi_1996, Buchm_ller_2005, Sakharov:1967dj, Kolb:1979qa, Khlebnikov:1988sr}.\\
Leptogenesis is most commonly studied in the context of thermal production of right-handed neutrinos, where these states are generated through scattering processes in the early Universe. Alternatively, in non-thermal scenarios \cite{Hahn_Woernle_2009, datta2023, datta2024, zhang2024, Chianese:2025mll,Alexander:2004us} the heavy neutrinos originate from the out-of-equilibrium decays of heavier particles such as the inflaton or the scalar field responsible for \( U(1)_{B-L} \) breaking. These non-thermal mechanisms can significantly lower the scale required for successful leptogenesis and are therefore of considerable interest. Note that the Sakharov condition of requiring a departure from equilibrium is achieved by the fact that these right handed neutrinos tend to be long lived. This opens the possibility that they are sufficiently long lived to cause a brief period of early matter domination which can leave an observable imprint on the gravitational wave background \cite{Datta:2025vyu}.

In both thermal and non-thermal cases, flavour effects play a crucial role: below temperatures of about \( 10^{12}\,\mathrm{GeV} \), the charged-lepton Yukawa interactions induce flavour decoherence, leading to distinct washout and \( CP \)-violating effects in each flavour. A consistent treatment of these flavour dynamics is therefore essential for accurately predicting the final baryon asymmetry \cite{Nardi_2006, Nardi:2006fx, Abada_2006, Antusch_2006, Blanchet_2007, DeSimone:2006nrs, Cirigliano_2010, Simone_2007, Racker_2012, Moffat_2018, baker2024hotleptogenesis, Ulysses, Ulysses2, blanchet2013leptogenesisheavyneutrinoflavours, Chianese_2024}. In this work, we consider flavoured thermal and non-thermal leptogenesis within a framework featuring a global and local \( U(1)_{B-L} \) symmetry.

\begin{figure}[t]
\centering
\begin{adjustbox}{max width=0.9\textwidth}
\begin{tikzpicture}[
    node distance=3.5cm and 3.0cm,
    every node/.style={font=\sffamily\bfseries, align=center},
    bubble/.style={draw, ellipse, thick, fill=blue!6, minimum width=3.5cm, minimum height=1.4cm, drop shadow},
    arrow/.style={-{Latex[length=3mm,width=2mm]}, thick},
    lab/.style={midway, sloped, above, font=\footnotesize\sffamily}
]

\node[bubble] (U1) {$\mathbf{U(1)_{B-L}}$\\symmetry};
\node[bubble, above right=of U1] (Maj) {Majorana Masses\\for RH Neutrinos};
\node[bubble, right=of Maj] (NuMass) {Neutrino\\ Masses};
\node[bubble, below=2.5cm of NuMass] (BAU) {Baryon Asymmetry\\of the\\ Universe (BAU)};
\node[bubble, below right=of U1] (CS) {Cosmic\\Strings};
\node[bubble, right=of CS] (GWB) {Gravitational\\Wave Background};

\draw[arrow] (U1) -- node[lab] {Symmetry Breaking} (Maj);
\draw[arrow] (Maj) -- node[lab] {Type-I seesaw} (NuMass);
\draw[arrow] (Maj) -- node[lab] {Leptogenesis} (BAU);
\draw[arrow] (U1) -- node[lab, below] {Symmetry Breaking} (CS);
\draw[arrow] (CS) -- node[lab] {Decays / Evolution} (GWB);

\end{tikzpicture}
\end{adjustbox}

\caption{\it
Schematic overview linking $U(1)_{B-L}$ symmetry breaking, neutrino mass generation, baryogenesis, cosmic strings and gravitational waves.  
The spontaneous breaking of the global $U(1)_{B-L}$ symmetry generates Majorana masses for right-handed neutrinos and simultaneously produces a network of cosmic strings.  
The heavy neutrinos yield light active neutrino masses via the Type-I seesaw and then also generates the Baryon Asymmetry of the Universe (BAU) through CP-violating decays.  
Meanwhile, the cosmic-string network evolves and sources a stochastic gravitational-wave background (GWB).}
\label{fig:schematic_overview}
\end{figure}
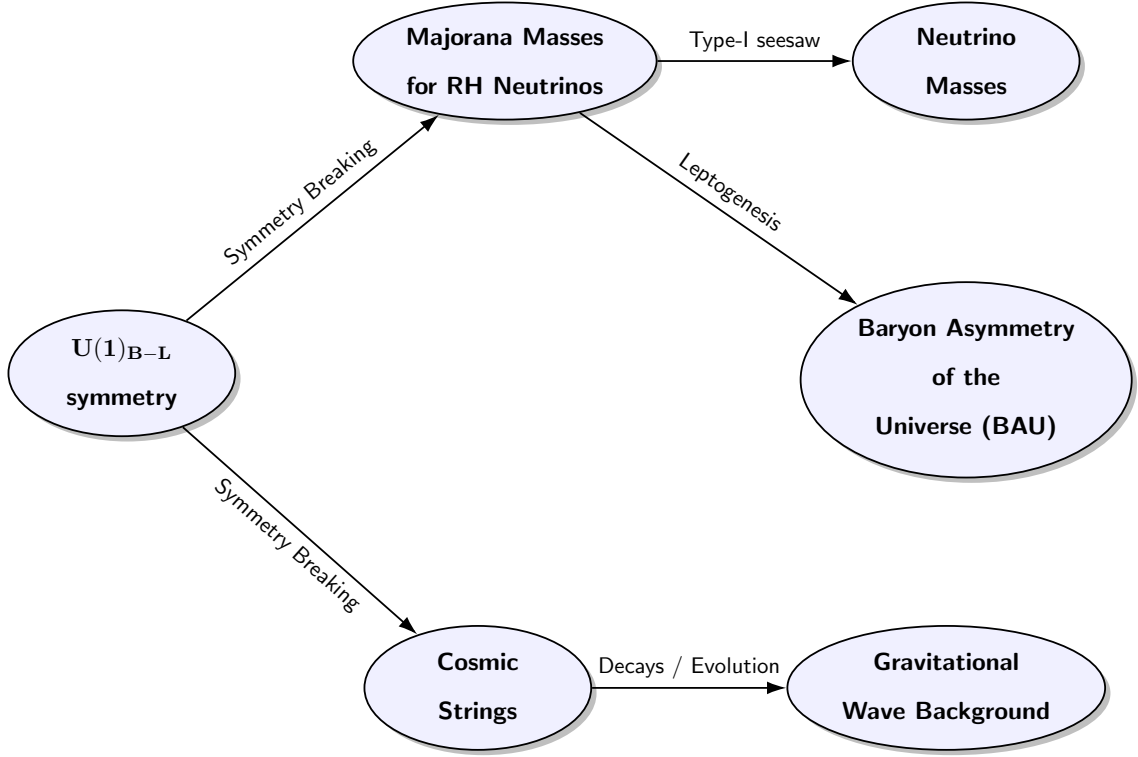

\begin{table}[h!]
\centering
\renewcommand{\arraystretch}{1.4}
\setlength{\tabcolsep}{12pt}
\begin{tabular}{|c|c|c|}
\hline
\textbf{GW from CS Testing the scale of} & \textbf{Gauged} $U(1)_{B-L}$ & \textbf{Global} $U(1)_{B-L}$\\
\hline
\textbf{Type I Seesaw} & Done in~\cite{Dror_2020}. & Done in~\cite{Fu_2023}. \\
\hline
\textbf{Vanilla Leptogenesis} & Done in~\cite{Dror_2020}. & \textbf{This Work.}\\
\hline
\textbf{Non-Thermal Leptogenesis} & \textbf{This Work.} & \textbf{This Work.}\\
\hline
\textbf{Near-resonant Leptogenesis} & \textbf{This Work} & \textbf{This Work.}\\
\hline
\textbf{Flavoured Leptogenesis} & Done in \cite{Chianese:2024gee}. & \textbf{This Work.}\\
\hline
\end{tabular}
\caption{\it Summary of existing and new work on Testing Type I Seesaw and Leptogenesis with gravitational waves from cosmic strings in Gauged and Global $U(1)_{B-L}$ models.}
\label{tab:Testing_Gauged_Global_Seesaw_Leptogenesis}
\end{table}
In addition to the thermal and non-thermal scenarios, we also consider near-resonant leptogenesis, in which the mass splitting between two right-handed neutrinos is comparable to their decay widths. Although fully resonant leptogenesis \cite{Pilaftsis_1997, Anisimov_2006, Garbrecht_2014, Garny_2013, Riotto_2007, Klaric_2021} can produce a maximal enhancement of the CP-asymmetry, it is often theoretically delicate and highly model-dependent. We therefore focus on the near-resonant regime, which provides a substantial enhancement of the asymmetry while remaining within a controlled uncontroversial framework. As we will demonstrate, this scenario considerably enlarges the parameter space for successful leptogenesis and can lower the required mass scale down to the $TeV$ range within our setup. This analysis has been done for considering testing some leptogenesis models with a local $U(1)_{B-L}$ and testing the type I seesaw for global but not yet leptogenesis, and in neither case for near-resonant leptogenesis or non-thermal leptogenesis. The purpose of this paper is two-fold: the first is to fill these gaps in the literature which we demonstrate in table \ref{tab:Testing_Gauged_Global_Seesaw_Leptogenesis}\footnote{Comparison between Dirac and Majorana leptogenesis probed via GW from CS was shown in Ref. \cite{Ghoshal:2025iho}.} and the second is to show that regions of the parameter space believed to be ruled out can be rescued by flavour effects. We also show a schematic for the general framework in Figure \ref{fig:schematic_overview}. 
\\
\textit{This paper is organised as follows:} In Section~\ref{sec: Cosmic Strings}, we review the formation of cosmic strings arising from the spontaneous breaking of a global or gauged $U(1)_{B-L}$ symmetry, and discuss the resulting gravitational-wave background and its testable parameter space in future experiments. In Section~\ref{sec:analytic leptogenesis}, we review the framework of thermal and non-thermal leptogenesis and derive analytic bounds on the right-handed neutrino mass required for successful baryogenesis, considering both thermal and non-thermal regimes. We also present the regime in which the analytic bounds remain valid. In Section~\ref{sec: numeric leptogenesis}, we extend the analysis to include flavour effects and near-resonant leptogenesis, providing numerical scans across the full parameter space of interest. The section ends with the detectable regions of successful leptogenesis by future gravitational wave experiments. Finally, we summarise our findings in Section~\ref{sec: conclusion}.

\section{Review of Gravitational Waves from Global and Local Cosmic Strings}
\label{sec: Cosmic Strings}
Gravitational Waves (GW) astronomy was initiated first with the detection of  gravitational waves from astrophysical sources by the LIGO collaboration in 2015~\cite{Abbott:2017xzu}. Upgrades with LIGO-Virgo-Kagra \cite{Aasi:2014mqd} consolidated this. Now with future GW detectors such as LISA~\cite{Audley:2017drz}, DECIGO~\cite{Yagi:2011wg} in space, or the 
Einstein Telescope~(ET)~\cite{Punturo:2010zz,Hild:2010id},
and Cosmic Explorer~(CE)~\cite{Evans:2016mbw} on the ground are expected to open up a new observational window of the early universe. Along with this the recent measurement of the stochastic GW background (SGWB) by the various pulsar timing array (PTA) collaboration actually gathered data of GW background ~\cite{Carilli:2004nx,Janssen:2014dka,Weltman:2018zrl,EPTA:2015qep,EPTA:2015gke,NANOGrav:2023gor,NANOGrav:2023hvm} investigating which primordial cosmology can be probed, see Ref. \cite{NANOGrav:2023hvm}. Unlike photons, primordial GWs are relics that were emitted from the early Universe, however they can propagate freely throughout the entire cosmic history of the Universe \cite{Allen:1996vm, Caprini:2018mtu,Simakachorn:2022yjy}.
Among various sources such as inflation, phase transitions, etc. one of the very suitable candidates that could help us to discern the pre-BBN cosmic expansion history of the universe is the detection of GW sourced by a network of cosmic strings. We describe below which such is the case. Cosmic strings are but one-dimensional objects produced by the spontaneous breaking of a $U(1)$ symmetry in the early universe~\cite{Nielsen:1973cs, Kibble:1976sj}, \footnote{In superstring theory, strings are fundamental objects and not topological defects~\cite{Copeland:2003bj, Dvali:2003zj, Polchinski:2004ia, Jackson:2004zg, Tye:2005fn}}.
A network of cosmic strings is formed whenever there is a  spontaneous breaking of a $U(1)$ symmetry, for instance,  global or gauged $U(1)_{\rm B-L}$ \cite{Kibble:1976sj, Vilenkin:2000jqa}. These are but topological defects in field theory solutions. As the universe evolves, long strings intercommute, leading to the production of closed string loops which afterwards oscillate and radiate gravitational radiation and Goldstone bosons (GB) for local and global cases respectively \cite{Vachaspati:1984gt, Allen:1991bk,Gouttenoire:2019kij, Auclair:2019wcv, Gouttenoire:2022gwi, Simakachorn:2022yjy}. The correlation length $L=\sqrt{\mu/\rho_\infty}$, of the strings related to the energy density $\rho_\infty$ in long strings, where $\mu$ is the string tension given by,
\begin{equation}
\mu = 2\pi n \, v_{\rm B-L}^2 \times \begin{cases}
1 ~ ~ ~ & {\rm for ~ local ~ strings},\\
\log(v_{\rm B-L} t)  ~ ~ ~ & {\rm for ~ global ~ strings}.
\end{cases}
\label{string_tension}
\end{equation} 
determines all the dynamics of such strings. Here $v_{\rm B-L}$ represents the vev of the scalar field which breaks the $U(1)_{\rm B-L}$ and $n$ is the winding number. We set $n=1$ since this is the only known stable configuration for gauged cosmic strings till now \cite{Laguna:1989hn}. For the global case we also assume it to be same for simplicity. 
It is important to note that for the scenario involving the global cosmic strings, we encounter a logarithmic dependence with respect to the VEV due to the presence of the massless Goldstone boson mode.  \\
The Goldstone bosons lead to typical long-range gradient energy as shown in Ref. \cite{Vilenkin:2000jqa}. The temperature at which CS network is formed can be understood as 
\begin{align}
T_\textrm{form} &\simeq v_{\rm B-L} \simeq 10^{11} \textrm{ GeV}\left(\frac{G\mu}{10^{-15}}\right)^{1/2},
\label{string_formation_cutoff}
\end{align}
where the second equality denotes the case for local strings.

Due to the emission of the GW or GB from the loops they suffer a decrease in their size following $l_i=\alpha t_i$ as,
\begin{align}
l(\tilde{t}) =  l_i - (\Gamma G \mu + \kappa) (\tilde{t} - t_i), \label{eq:length_shrink}
\end{align}
where $\Gamma\simeq 50$ \cite{Vilenkin:2000jqa, Vachaspati:1984gt}, $\alpha\simeq0.1$ \cite{Blanco-Pillado:2013qja,Blanco-Pillado:2017oxo}, $G$ is the Newton's gravitational constant, and $t_i$ is the time of loop formation. The shrinking rate is affected by two contributions, $\Gamma G\mu$, associated with GW emission, and $\kappa$, associated with Goldstone production. For local strings, $\kappa = 0$, and loop decay is dominated by gravitational radiation. In contrast, global cosmic strings emit predominantly into Goldstone bosons, and decays with an efficiency of
$\kappa = \frac{\Gamma_{\rm Gold}}{2\pi} \log(v_{\rm B-L} t) \gg \Gamma G \mu$ ,
where  $\Gamma_{\rm Gold} \simeq 65$ \cite{Vilenkin:1986ku, Gorghetto:2021fsn}. For discussions regarding the production of gravitational radiation or simply GW from cosmic strings, see reviews in Refs. \cite{Gouttenoire:2019kij, Auclair:2019wcv, Gouttenoire:2022gwi, Simakachorn:2022yjy}. 

 After its formation, the cosmic string network interacts strongly with the thermal plasma leading to some kind of damped motion. Afterwards this the strings oscillate.
The Hubble expansion and cosmic string loop formation, respectively increase and decrease the fractional energy budget abundance $\Omega_{\rm CS} \equiv \rho_{\rm CS} / \rho_{\rm tot}$. This happens in a manner the the cosmic string network reaches a stable state. This configuration is known as the \emph{scaling regime} where the abundance $\Omega_{\rm CS}$ remains constant over time even with the evolution of the universe \cite{Albrecht:1984xv, Bennett:1987vf, Allen:1990tv,Martins:2000cs, Figueroa:2012kw, Martins:2016ois}. During this scaling regime, CS redshifts in a  same way like that of the background, e.g., $\rho_{\rm CS} \propto a^{-4}$ during radiation-domination (RD) and  $\rho_{\rm CS} \propto a^{-3}$ during the matter-domination (MD) epoch. This means that the equation of state of the CS network tracks the equation of state of the expanding background \cite{Vachaspati:1984gt, Allen:1991bk,Vilenkin:2000jqa} .



In spite of these two competing dynamics lies an attractor solution known as the scaling regime, in which the characteristic length $L$ scales as cosmic time $t$. In this scaling regime, for a constant string tension, the energy density evolves as $\rho_\infty \propto t^{-2}$. Consequently, the cosmic string network becomes the same as any cosmological background energy density $\rho_{\rm bkg} \propto t^{-2}$ with the constant of proportionality given by the small quantity $G\sim M_{\rm Pl}^{-2}$, where $G$ is the Newton constant. Unlike topological defects like domain wall network or other cosmological relics like dark matter, axions this particular scaling behaviour of the cosmic string network makes sure that there is cosmic sting domination over the Universe's energy density budget. We remark here the difference between local and global cosmic strings now: the local string loops decay into gravitons as GWs after $0.001/G\mu \gg 1$ Hubble times, while global string loops decay into Goldstone radiation modes in less than one Hubble time as usual.

In a nutshell, first, the string loops are created at say time $t_i$ with string loop size parameter denoted by $\alpha$. The rate of loop formation rate being be expressed as \cite{Vilenkin:2000jqa}:

\begin{align}
\frac{dn_{\rm loop}}{dt_i} = 0.1 \frac{C_{\rm eff}(t_i)}{\alpha t_i^4}.
\end{align}
where the factor $0.1$ means that $90\%$ of the loop population consists of small, highly-boosted loops. Consequently, these are usually red-shifted away without contributing much to the GW radiation in any significant manner \cite{Blanco-Pillado:2013qja}. The loop formation efficiency factor $C_{\rm eff}$, which expresses the decay rate at which local cosmic strings attain the asymptotic values or scaling solutions is given by $C_{\rm eff} \simeq 0.39$ and $5.4$ during the matter-dominated (MD) and radiation-dominated (RD) era respectively \cite{Gouttenoire:2019kij}. 
For global strings, besides the loop formation, the long strings also emit energy into light Goldstone boson particle production, due to which the loop production efficiency becomes logarithmically suppressed with time. Analytically, this can be somewhat understood as  $C_{\rm eff} \sim \mathcal{O}(1)$ for all cosmological epochs \cite{Chang:2019mza, Gouttenoire:2019kij,Chang:2021afa,Ramberg:2019dgi}. 
 In order to accurately determine $C_{\rm eff}(t)$ one solves the velocity-dependent one-scale (VOS) equations which govern the evolution of the string network. However, in present analysis, such is beyond the scope and we employ some simplistic scaling solutions instead. These solutions effectively capture the essential qualitative features of the model that we need to demonstrate for the complementarity studies that we allude to, in this paper.

Following numerical simulations which suggest that the GW is emiited predominantly by loops of the largest size \cite{Blanco-Pillado:2013qja} one takes a monochromatic probability distribution of loop sizes
\begin{align}
\mathcal{P}_{\rm loop}(\alpha) = \delta(\alpha - 0.1)
\end{align}
from the time of their formation at $\tilde t > t_i$. From there-on loops of length $l(\tilde{t})$ oscillate and radiate a discrete spectrum of GWs with frequencies given by the relation
\begin{equation}
\label{eq:emitted_frequency}
\tilde{f} = 2k/l(\tilde{t}), \qquad  k \in \mathbb{Z}^+.
\end{equation}
The present-day red-shifted frequency is therefore set by $f=\tilde{f} a(\tilde{t})/a_0$.

Now, for each Fourier modes, GW emission power is given by the relation
\begin{align}
P_{\rm GW}^{(k)} = \Gamma^{(k)} G\mu^2, \qquad \text{with} \quad \Gamma^{(k)} = \frac{\Gamma k^{-{\delta}}}{\sum_{p=1}^\infty p^{-\delta}},
\label{eq:power_emisssion_GW_strings}
\end{align}
where $\Gamma = 50$ for local cosmic strings \cite{Blanco-Pillado:2017oxo} and global global strings \cite{Gorghetto:2021fsn}. 
The value of $\delta$ captures whether high Fourier modes are dominated by cusps (for which $\delta = 4/3$), kinks (for which $\delta =5/3$), or kink-kink collisions  (for which $\delta=2$) \cite{Olmez:2010bi}. Typically it is assumed that the small-scale case to be dominated by cusps. As the loops continuously lose energy via the emission of GWs or via Goldstone boson production, 
the length $l$ of the loops keeps on shrinking as 
\begin{align}
l(\tilde{t}) =  \alpha t_i - (\Gamma G \mu + \kappa) (\tilde{t} - t_i). \label{eq:length_shrink}
\end{align}
where $\Gamma G \mu$ and $\kappa$ are the shrinking rates of the emission into GW background and particle productions, respectively.
As stated above since the local-string loops primarily decay via gravitational radiation for which one takes $\kappa = 0$. While the global string loops predominantly decays away via producing Goldstone bosons efficiently, which is essentially characterised by $\kappa = \Gamma_{\rm Gold}/2 \pi \log(v_{\rm B-L} t) \gg \Gamma G\mu$, where $\Gamma_{\rm Gold} \simeq 65 $ \cite{Vilenkin:1986ku}. 
At the end of the day, the total GW energy emitted by a loop can be decomposed into several harmonics with instantaneous frequencies
\begin{equation}\label{eq:emitted_frequency}
    f_k = \frac{2k}{\ell_k} = \frac{a(t_0)}{a(t)} f,
\end{equation}
where $k = 1, 2, 3, \dots, k_{\rm max}$, $f$ is the frequency observed today at $t_0$, and $a(t)$ is the FRW scale factor. The total GW energy density is estimated by summing over all the $k$ Fourier modes giving us
\begin{align}
\Omega_{\rm GW}(f) &=\sum_k\frac{1}{\rho_c}\cdot\frac{2k}{f}\cdot\frac{\mathcal{F}_\alpha \,\Gamma^{(k)}G\mu^2}{\alpha(\alpha+\Gamma G \mu + \kappa)} \times \nonumber\\
&\hspace{3em}  \int^{t_0}_{t_{\rm osc}}d\tilde{t} \, \frac{C_{\rm{eff}}(t_i)}{t_i^4}\left[\frac{a(\tilde{t})}{a(t_0)}\right]^5\left[\frac{a(t_i)}{a(\tilde{t})}\right]^3\Theta\left(t_i-\frac{l_*}{\alpha}\right)\Theta(t_i-t_{\rm osc}).
	\label{eq:master_eq_ready_to_use}
	\end{align}
Here, $\rho_c$ denotes the critical energy density of the universe as usual, $\mathcal{F}_\alpha \simeq 0.1$ is an efficiency factor,  and $C_{\rm eff}(t_i)$ is the cosmic string loop formation efficiency, which gives us an estimate from the velocity-dependent one-scale model \cite{Martins:1996jp, Martins:2000cs, Sousa:2013aaa, Auclair:2019wcv, Sousa:2020sxs}. The integral in Eq. \eqref{eq:master_eq_ready_to_use} is associated with two Heaviside functions, $\Theta\left(t_i-\frac{l_*}{\alpha}\right)\Theta(t_i-t_{\rm osc})$, which regulate
and impose high-frequency cut-off at $f_*$, beyond
which the GW spectrum exhibits a slope $f^{-1/3}$ when summed over a very large number of Fourier modes. The quantity $t_{\rm osc}=\text{Max}\left[t_{\rm form},\,t_{\rm fric} \right]$ marks the time period at which the motion of the string network stops to be friction-dominated. The parameter $l_*$ represents a critical loop length above which GW emission always becomes dominant over the particle production. As expected, Eq.~\eqref{eq:master_eq_ready_to_use} is applicable to both local and global strings, given that  Eq.~\eqref{eq:emitted_frequency} is used in association with an appropriate choice of $\kappa$.

At higher frequencies, within standard cosmological evolution, the GW spectrum from local strings gives approximately scale-invariant, the ampltiude of the formed GW
\begin{equation}
    \Omega_{\rm std}^{\rm local} h^2 \simeq 15 \pi\Omega_r h^2 \, \Delta_T\, C_{\rm eff}^{\rm rad,l} \mathcal{F}_\alpha\left(\frac{\alpha G \mu}{\Gamma}\right)^{1/2},
\end{equation}
where $\Omega_{r}h^2 \simeq 4.2 \times 10^{-5}$\cite{ParticleDataGroup:2020ssz}. Small deviations from the flatness of the potential could arise due to variations in the number of relativistic degrees of freedom, captured in \cite{Gouttenoire:2019kij} by
\begin{align}
\label{eq:Delta_R}
\Delta_T \equiv \left( \frac{g_*(T)}{g_*(T_0)}\right)\left(\frac{g_{*s}(T_0)}{g_{*s}(T)} \right)^{4/3} .
\end{align}
In contrast, the high-frequency part of the GW spectrum from global strings is quite suppressed. This can be approximately written as \cite{Gouttenoire:2019kij}
\begin{equation}
    \Omega_{\rm std}^{\rm global} h^2 \sim  90\Omega_r h^2 \, \Delta_T\,C_{\rm eff}^{\rm rad,g} \mathcal{F}_\alpha\left(\frac{\Gamma}{\Gamma_{\rm gold}}\right) \left(\frac{v_{\rm B-L}}{M_{\rm pl}}\right)^{4} \log^3\left( v_{\rm B-L} \tilde{t}_{\rm M}\right),
\end{equation}
where the time of maximum emission is
\begin{equation}
    \tilde{t}_M =\frac{1}{t_0} \frac{4}{\alpha^2} \left( \frac{1}{f}\right)^2 \left( \frac{\alpha+\Gamma G \mu +\kappa}{\Gamma G \mu +\kappa}\right)^2.
\end{equation}
A detailed comparison between the GW signatures emitted from local and global strings can be found in Refs.~\cite{Gouttenoire:2019kij,Ghoshal:2023sfa}.\\

\begin{figure}[H]
\centering
\includegraphics[height=6cm,width=0.49\linewidth]{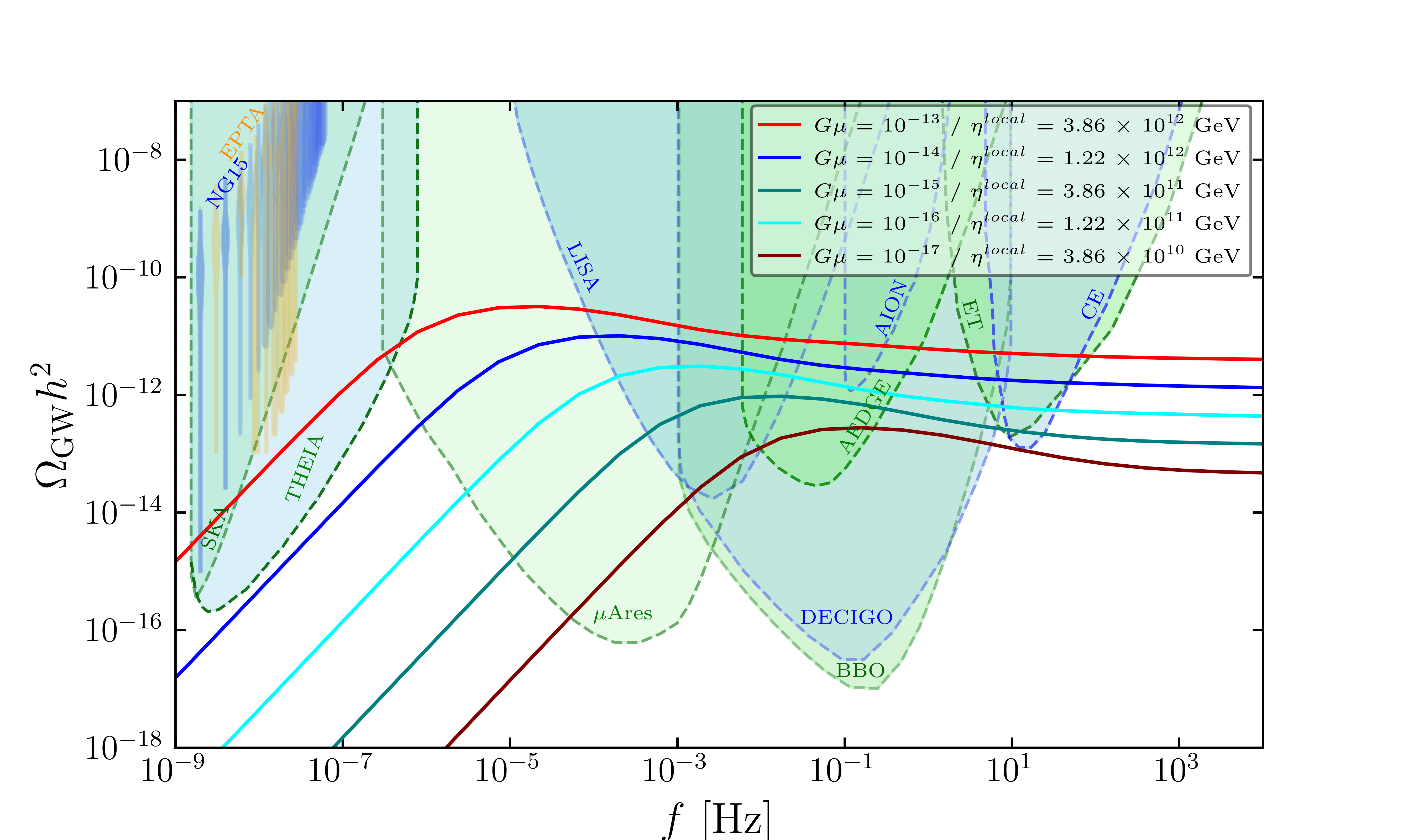}
\includegraphics[height=6cm,width=0.49\linewidth]{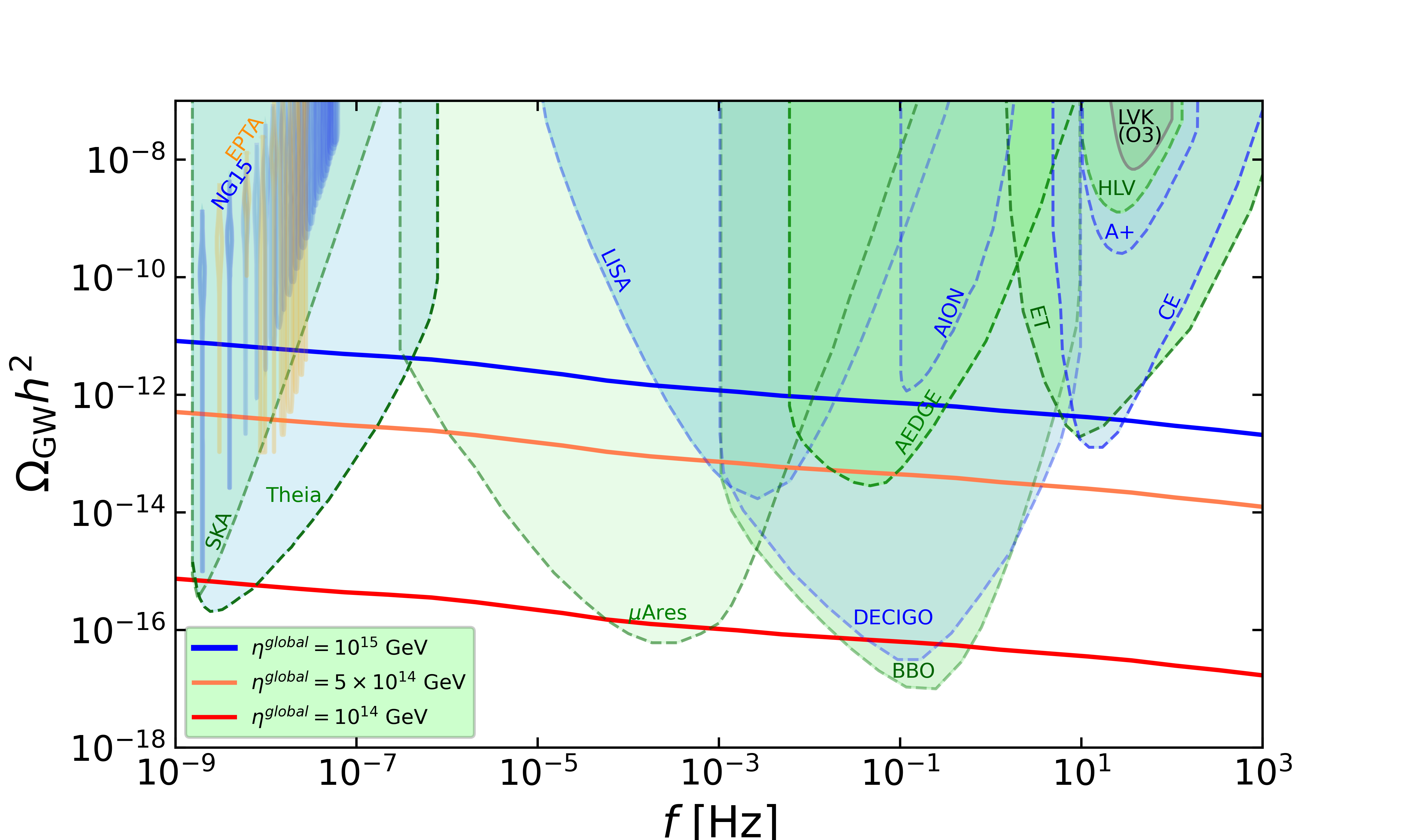} 
\caption{\it The GW spectral shapes for local strings (\textbf{left panel}) and global strings (\textbf{right panel}) are
shown. $\eta$ denotes vev $v_{\rm B-L}$ here.}
    \label{fig:CS benchmark}
\end{figure}

\subsection{Present constraints on cosmic strings}

\textbf{Local strings:} A SGWB detected by pulsar timing arrays NANOGrav \cite{NANOGrav:2020bcs}, EPTA \cite{Chen:2021rqp}, PPTA \cite{Goncharov:2021oub}, and IPTA \cite{Antoniadis:2022pcn}. This signal has a favorable interpretation as an SGWB from cosmic strings with tension $G\mu \sim 8 \times 10^{-11}$ \cite{Ellis:2020ena,Blasi:2020mfx}.

\textbf{Global strings:}
Since Global cosmic strings efficiently decay into massless Goldstone particles they contribute to the number $N_{\rm eff}$ of effective relativistic degrees of freedom or dark radiation. The precise constraint depends upon the abundance of Goldstone particles from strings which is still a debatable result in the community, see Refs.  \cite{Hindmarsh:2019csc,Hindmarsh:2021vih,Buschmann:2019icd,Buschmann:2021sdq} and its comparative results with Refs. \cite{Gorghetto:2018myk,Gorghetto:2020qws,Gorghetto:2021fsn} for recent studies on this topic. However in our analysis we simply take the upper bound $\eta \lesssim 3.5 \cdot 10^{15} \, {\rm GeV}$ derived in Ref.~\cite{Chang:2021afa}.  

The constraint on the scale of inflation $H_{\rm inf}\lesssim 3\times 10^{13}~\rm GeV$ from CMB missions \cite{BICEP:2021xfz} translates to the maximum temperature of the universe $T_{\rm max} \lesssim 4 \times 10^{15}$~GeV. For the cosmic string network to form, the string scale $\eta$ should be smaller than the maximum temperature $\eta \lesssim 4 \times 10^{15}$~GeV, up to $\mathcal{O}(1)$ factors depending upon various model-dependent parameters.

For $\eta \gtrsim 10^{15} \, {\rm GeV}$, GW from global strings extend to $f \lesssim 10^{-14} \, {\rm Hz}$ which could leave signature in CMB B-mode polarization experiments, such e.g. Ref.~\cite{BICEP:2021xfz}, however this constraint is not very strict since this emmision occurs after photon decoupling in CMB, see
Fig.~8 of Ref.~\cite{Chang:2021afa} for some involved details.

\medskip

\subsection{Gravitational Wave Detectors} 
In the GW spectrum plots, in presented in Fig. \ref{fig:CS benchmark}, we display the power-law integrated sensitivity curves for a myriad of ongoing and future GW missions. Usually they can be grouped as: 
\begin{itemize}
    \item \textbf{Ground-based interferometers:} These detectors, such as \textsc{LIGO}/\textsc{VIRGO} \cite{LIGOScientific:2016aoc,LIGOScientific:2016sjg,LIGOScientific:2017bnn,LIGOScientific:2017vox,LIGOScientific:2017ycc,LIGOScientific:2017vwq}, a\textsc{LIGO}/a\textsc{VIRGO} \cite{LIGOScientific:2014pky,VIRGO:2014yos,LIGOScientific:2019lzm}, \textsc{AION} \cite{Badurina:2021rgt,Graham:2016plp,Graham:2017pmn,Badurina:2019hst}, \textsc{Einstein Telescope (ET)} \cite{Punturo:2010zz,Hild:2010id}, and \textsc{Cosmic Explorer (CE)} \cite{LIGOScientific:2016wof,Reitze:2019iox}, which use interferometer-based techniques on the Earth's surface to detect gravitational waves.
    
    \item \textbf{Space-based interferometers:} Space-based detectors like \textsc{LISA} \cite{Baker:2019nia}, \textsc{DECIGO}, \textsc{U-DECIGO} \cite{Seto:2001qf,Yagi:2011wg}, \textsc{AEDGE} \cite{AEDGE:2019nxb,Badurina:2021rgt}, and \textsc{$\mu$-ARES} \cite{Sesana:2019vho} are designed to detect gravitational waves from space, also by interferometric techniques offering different advantages over ground-based counterparts.
    
    \item \textbf{Recasts of star surveys:} Monitoring of star surveys like \textsc{GAIA}/\textsc{THEIA} \cite{Garcia-Bellido:2021zgu} utilize astrometric data from stars. They can indirectly infer the presence of gravitational wave signals.
    
    \item \textbf{Pulsar timing arrays (PTA):} PTA experiments like \textsc{SKA} \cite{Carilli:2004nx,Janssen:2014dka,Weltman:2018zrl}, \textsc{EPTA} \cite{EPTA:2015qep,EPTA:2015gke}, and \textsc{NANOGRAV} \cite{NANOGRAV:2018hou,Aggarwal:2018mgp,NANOGrav:2020bcs} use precise timing periodicity measurements of pulsars to infer the presence of gravitational wave signatures.
\end{itemize}

\subsection{Signal-to-noise ratio (SNR)}
\label{SNR}
 Gravitational Wave Interferometers deals with the measurements of displacements of detector arms via the dimensionless strain-noise $h_\text{GW}(f)$ which is related to the GW amplitude. This quantity is usually converted into the corresponding GW energy density in the form \cite{Garcia-Bellido:2021zgu}
\begin{align}
    \Omega_\text{exp}(f) h^2 = \frac{2\pi^2 f^2}{3 H_0^2} h_\text{GW}(f)^2 h^2,
\end{align}
where $H_0 = h\times 100 \;\text{(km/s)}/\text{Mpc}$ is the Hubble cosmic expansion rate as observed today. 
The signal-to-noise ratio (SNR) then for any projected experimental sensitivity $\Omega_\text{exp}(f)h^2$ is estimated in order to assess the detection prospects of the primordial GW background originating from the global and local cosmic string background following the prescription as given in details in Refs.~\cite{Thrane:2013oya,Caprini:2015zlo}
\begin{align}
     \text{SNR}\equiv \sqrt{\tau \int_{f_\text{min}}^{f_\text{max}} \text{d}f \left(\frac{ \Omega_\text{GW}(f) h^2}{\Omega_\text{exp}(f) h^2}\right)^2 } \label{eq:SNR},
\end{align}
where $h=0.7$ and  $\tau = 4\; \text{years}$ (for all interferometer based GW detectors) and $\tau = 20\; \text{years}$ (for PTA based GW detectors) is the observational time. We consider $\text{SNR}\geq 10$ always as the detection threshold for each individual GW detector in our present analysis.

\medskip

\section{Leptogenesis with a $U(1)_{B-L}$ Extension}
\label{sec:analytic leptogenesis}

The origin of the matter–antimatter asymmetry remains one of the central open questions in cosmology and particle physics. The baryon asymmetry of the Universe has been precisely determined through independent measurements from the cosmic microwave background~\cite{Planck2018} and Big Bang nucleosynthesis~\cite{ParticleDataGroup:2020ssz} 
\begin{equation}
    Y_B=\frac{n_B}{s}= 8.87\times 10^{-11}\ .
    \label{eq:BAU}
\end{equation}
Leptogenesis~\cite{Fukugita:1986hr} provides a compelling framework for explaining the matter–antimatter asymmetry of the Universe. In this mechanism, CP-violating decays of heavy right-handed neutrinos satisfying the Sakharov conditions~\cite{Sakharov:1967dj} generate a net lepton asymmetry, which is subsequently partially converted into a baryon asymmetry through electroweak sphaleron processes~\cite{Luty:1992un, Giudice_2004, Covi_1996, Buchm_ller_2005}. In this section we review the Type-I seesaw framework and analytically solve the relevant Boltzmann equations in the weak-washout regime. This allows us to derive a lower bound on the right-handed neutrino mass required for successful leptogenesis and to specify precisely when this bound is applicable. We also determine the additional conditions needed to ensure the consistency of our analysis by ensuring the universe remains radiation-dominated throughout the production and decay era which is integral to calculating the baryon asymmetry and the gravitational wave spectrum. Finally, we outline the circumstances under which the bound fails and flavour effects must be taken into account.
\subsection{The Type I Seesaw Mechanism}
\label{sec:domination}
The seesaw mechanism provides a natural explanation for the smallness of active neutrino masses by introducing heavy states at high energy scales: fermion singlets (right-handed neutrinos) in the type-I seesaw \cite{Minkowski:1977sc, Gell-Mann:1979vob, Yanagida:1979as, Mohapatra:1980yp}, scalar triplets in the type-II seesaw \cite{Magg:1980ut, Wetterich:1981bx, Schechter:1980gr}, and fermion triplets in the type-III seesaw \cite{Foot:1988aq, Ma_1998, Ma_2002}. In each case, the light neutrino masses emerge inversely proportional to the masses of these heavy states,
\begin{equation}
    m_{\nu}\propto\frac{v_H^2}{M}
\end{equation}
where $M$ is the mass of the heavy seesaw state and $v_H$ is the Higgs vev. For this work we consider the type I seesaw. A large value of $M$ naturally accounts for the smallness of active neutrino masses. Moreover, the presence of a $U(1)_{B-L}$ symmetry offers a natural origin for the heaviness of $M$: the masses of the heavy seesaw states are linked to the scale of $B-L$ symmetry breaking, $v_{B-L}$. Consequently, if the $B-L$ breaking scale is high, the corresponding seesaw masses are naturally large as well. The particle content for the type I seesaw with a local or global $U(1)_{B-L}$ symmetry is shown in Table \ref{tab:BL-table}.
\begin{table}[h!]
\centering
\begin{tabular}{c|c c c c}
\hline
Field & $SU(3)_c\ $ & $SU(2)_L\ $ & $U(1)_Y\ $ & $U(1)_{B-L}$ \\
\hline
$q^i_L$     & $\mathbf{3}$ & $\mathbf{2}$ & $+1/6$ & $+1/3$ \\
$u^i_R$     & $\mathbf{3}$ & $\mathbf{1}$ & $+2/3$ & $+1/3$ \\
$d^i_R$     & $\mathbf{3}$ & $\mathbf{1}$ & $-1/3$ & $+1/3$ \\
$\ell^i_L$  & $\mathbf{1}$ & $\mathbf{2}$ & $-1/2$ & $-1$ \\
$e^i_R$     & $\mathbf{1}$ & $\mathbf{1}$ & $-1$   & $-1$ \\
$H$         & $\mathbf{1}$ & $\mathbf{2}$ & $-1/2$ & $0$ \\
\hline
$\Phi$      & $\mathbf{1}$ & $\mathbf{1}$ & $0$    & $+2$ \\
$N^i$     & $\mathbf{1}$ & $\mathbf{1}$ & $0$   & $-1$ \\
$Z'$    & $\mathbf{1}$ & $\mathbf{1}$ & $0$   & $0$ \\
\hline
\end{tabular}
\caption{\it Particle content of the $B-L$ extended Seesaw Model. The first block lists the Standard Model particles. The second block shows the additional scalar $\Phi$ required to break $U(1)_{B-L}$ and three heavy right-handed neutrinos, $N^i$. For a local symmetry the associated $Z'$ gauge boson is also present unlike when $U(1)_{B-L}$ is global.}
\label{tab:BL-table}
\end{table}

In the global type-I seesaw framework, the Lagrangian benefits from extra Yukawa interactions 
\begin{equation}
    \mathcal{L} \;\supset\; 
    -Y_{\alpha i}\, \overline{L_L^\alpha} \, \tilde{H} \, N_i 
    - \tfrac{1}{2}\, y_i\, \phi\, \overline{N_i^c}\, N_i 
    \;+\; \text{h.c.},
\end{equation}
where $Y_{\alpha i}$ is the neutrino Yukawa coupling matrix where the indices denote flavour, and $\tilde{H} = i \sigma_2 H^*$ is the conjugate Higgs doublet.
The scalar field $\phi$, carrying lepton number $L = -2$, is responsible for spontaneously breaking the global $U(1)_{B-L}$ symmetry. Once $\phi$ acquires a vacuum expectation value $\langle \phi \rangle = v_{B-L}$, the heavy Majorana masses 
$M_i = y_i v_{B-L}$ are generated, giving rise to the standard type-I seesaw mechanism for light neutrino masses. After electroweak symmetry breaking, 
the Dirac mass matrix is $m_D = Y v_h$. In the seesaw limit $M \gg m_D$, naturally achieved for a high $B-L$ symmetry breaking scale, the lightness of the active neutrino masses follows as a direct consequence.
\begin{equation}
    m_\nu \;=\; - m_D M^{-1} m_D^T
    \label{eq:seesaw}
\end{equation}
The general solution for the yukawa matrix $Y$ consistent with (\ref{eq:seesaw}) can be written as
\begin{equation}
    Y \;=\; \frac{1}{v_H}\, U \, \sqrt{m_\nu} \, R \, \sqrt{M}
    \label{eq:CI}
\end{equation}
where $U$ is the PMNS matrix, $m_\nu$ the diagonal light neutrino mass matrix, $M$ the heavy neutrino mass matrix, and $R$ is a complex orthogonal matrix. This parametrisation, known as the Casas-Ibarra form, ensures that the seesaw relation is automatically satisfied. The decay rate of a right–handed neutrino can then be written in a compact form
\begin{equation}
    \Gamma_{N_i} = \frac{(Y^\dagger Y)_{ii}}{8\pi}\,M_i=\frac{M_i^2\tilde m_i}{8\pi v_H^2}\ .
\end{equation}
This defines the effective neutrino mass
\begin{equation}
    \tilde m_i \equiv \sum_j m_j |R_{j i}|^2,
\end{equation}
where $R_{ji}$ are elements of the complex orthogonal matrix in the casas-ibarra parameterisation \ref{eq:CI}. When flavour must be tracked, it is convenient to introduce
\begin{equation}
    \tilde m_{i\alpha} \equiv \frac{v_H^2}{M_i}\,|Y_{\alpha i}|^2\ .
\end{equation}
By construction, these satisfy $\sum_\alpha \tilde m_{1\alpha} = \tilde m_1$. For flavourless thermal leptogenesis the bounds on the baryon asymmetry is uniquely determined by the two parameters $\tilde m_1$ and $M_1$. For flavoured leptogenesis we must consider the flavour projected effective masses as well and optimise them to give the maximum baryon asymmetry to find the true bounds. We note that leptogenesis can also proceed via the Type-II~\cite{Ma_2002, Hambye_2001,Joshipura_2001,Hambye_2004,Guo_2004,Antusch_2004,Antusch_2006,Gu_2006, McDonald_2008} and Type-III~\cite{Hambye_2004,Brahmachari_2001,Aristizabal_Sierra_2010,Blanchet_2008} seesaw mechanisms. In this work, however, we focus on the standard Type-I framework, which remains the most widely studied and phenomenologically established minimal realisation of leptogenesis.
\subsection{Ensuring Radiation Domination}
\label{sec: rad_dom}
Standard leptogenesis, and our analysis, assumes a radiation-dominated Universe. It is therefore essential to verify that this assumption remains valid throughout the production and decay of the heavy neutrinos. If the decays occur too late, the heavy states can temporarily dominate the energy density, invalidating our treatment. Consider a particle species $N$ of mass $M$ that decouples while still relativistic. For a relativistic species, both the number and entropy densities scale as temperature cubed, 
yielding a temperature-independent constant thermal yield
\begin{equation}
    n(T) = \frac{\zeta(3)}{\pi^2}\, g\, T^3, 
    \qquad 
    s(T) = \frac{2\pi^2}{45}\, g_*(T)\,T^3, 
    \qquad 
    Y \;\equiv\; \frac{n}{s} = 0.0026\,g.
\end{equation}
Once the temperature significantly drops, $T \ll M$, the species becomes non-relativistic and its energy density can be approximated by $\rho_N = M\,Y\,s(T)$. 
In this regime, the average energy has a subdominant kinetic component and can be approximated by the mass, $\langle E \rangle \simeq M + \mathcal{O}(T)$. 
Equating the energy densities of $N$ and radiation, $\rho_N = \rho_R$, defines the temperature at which the species begins to dominate \cite{Datta:2025vyu}:
\begin{equation}
    T_{\rm dom} = \frac{4}{3}\, M\,Y \simeq 0.37\,\frac{g}{g_*}\,M.
\end{equation}
This relation between $T_{dom}$ and the yield is also valid for non-thermal abundances. For right-handed neutrinos and a scalar field $\phi$, using the thermal yields, one finds
\begin{equation}
\begin{split}
    Y_N^i &= 3.9\times 10^{-3},\qquad\quad T_{\rm dom}^N = 0.52\%\,M,\\
    Y_\phi^i &= 5.2\times 10^{-3},\qquad\quad T_{\rm dom}^\phi = 0.69\%\,M,\\
    Y_{\rm tot} &= Y_N^i + 2Y_\phi^i = 1.43\times 10^{-2},\qquad 
    T_{\rm dom}^{\rm tot} = 1.9\%\,M.
\end{split}
\label{eq:Tdom}
\end{equation}
If $\phi$ decays before $N$, it can enhance the $N$ abundance by up to $2Y_\phi^i$, requiring a reassessment of the decay dynamics. The relevant decay and Hubble rates are
\begin{equation}
    \Gamma_N = \frac{M^2\tilde{m}}{8\pi}, 
    \qquad 
    \Gamma_\phi = \frac{y^2 M_\phi}{8\pi}, 
    \qquad 
    H(T) = 1.66\,\sqrt{g_*}\,\frac{T^2}{M_{\rm Pl}}.
\end{equation}
To ensure that decays occur before the onset of early matter domination, via either species we impose $\Gamma_N > H(T_{\rm dom}^{tot})$ and $\Gamma_\phi>H(T_{\rm dom}^{\phi})$, for thermal abundences this leads to the simple conditions
\begin{equation}
    \tilde{m}_i > 3.9\times 10^{-7}, 
    \qquad 
    y_i > 4.1\times 10^{-11}\sqrt{\frac{M_\phi}{\mathrm{GeV}}}.
    \label{eq: MD bounds}
\end{equation}
Given that $\phi$ may have a large mass, a period of early matter domination is therefore quite plausible, for example at $M_{\phi}=10^{13}$, the yukawa only needs to be of order $10^{-4}$ for early matter domination to occur. The possibility of right-handed neutrino domination has been discussed previously in \cite{Datta:2025vyu}, the analogous case of $\phi$ domination have been discussed in \cite{Blasi_2020, Datta:2022tab, Borboruah:2024eha}.

\subsection{The CP asymmetry}
\label{sec: CP asymmetry}
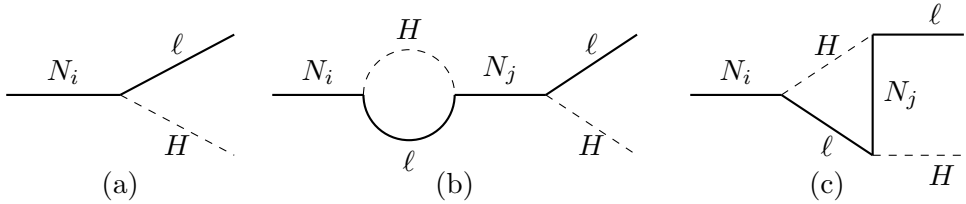
\begin{figure}[h!]
\centering
\begin{tikzpicture}[scale=1, baseline=(current bounding box.center)]

  \coordinate (Ni1) at (0,0);
  \coordinate (v1) at (1.5,0);
  \coordinate (ell1) at (3,0.8);
  \coordinate (phi1) at (3,-0.8);
  \draw[thick] (Ni1) -- (v1) node[midway, above] {\( N_i \)};
  \draw[thick] (v1) -- (ell1) node[midway, above ] {\( \ell \)};
  \draw[dashed] (v1) -- (phi1) node[midway, below ] {\( H \)};
  \node at (1.5, -1.2) {(a)};

  \begin{scope}[xshift=3.5cm]
    \coordinate (Ni2) at (0,0);
    \coordinate (loopL) at (1.2,0);
    \coordinate (loopR) at (2.4,0);
    \coordinate (v2) at (3.6,0);
    \coordinate (ell2) at (4.8,0.8);
    \coordinate (phi2) at (4.8,-0.8);
    \draw[thick] (Ni2) -- (loopL) node[midway, above] {\( N_i \)};
    \draw[thick] (loopR) -- (v2) node[midway, above] {\( N_j \)};
    \draw[thick] (v2) -- (ell2) node[midway, above ] {\( \ell \)};
    \draw[dashed] (v2) -- (phi2) node[midway, below ] {\( H \)};
    \draw[dashed] (loopL) arc[start angle=180, end angle=0, x radius=0.6cm, y radius=0.6cm] node[midway, above] {\( H \)};
    \draw[thick] (loopL) arc[start angle=-180, end angle=0, x radius=0.6cm, y radius=0.6cm] node[midway, below] {\( \ell \)};
    \node at (2.4, -1.2) {(b)};
  \end{scope}

  \begin{scope}[xshift=9cm]
    \coordinate (Ni3) at (0,0);
    \coordinate (vtx) at (1.2,0);
    \coordinate (Nj) at (2.4,0);
    \coordinate (ellOut) at (3.6,0.8);
    \coordinate (phiOut) at (3.6,-0.8);
    \coordinate (ellLoop) at (2.4,-0.8);
    \coordinate (phiLoop) at (2.4,0.8);
    \draw[thick] (Ni3) -- (vtx) node[midway, above] {\( N_i \)};
    \draw[thick] (vtx) -- (ellLoop) node[midway, below] {\( \ell \)};
    \draw[dashed] (vtx) -- (phiLoop) node[midway, above] {\( H \)};
    \draw[thick] (ellLoop) -- (phiLoop) node[midway, right] {\( N_j \)};
    \draw[thick] (phiLoop) -- (ellOut) node[midway, above right] {\( \ell \)};
    \draw[dashed] (ellLoop) -- (phiOut) node[midway, below right] {\( H \)};
    \node at (1.8, -1.2) {(c)};
  \end{scope}

\end{tikzpicture}
\caption{\it Feynman diagrams contributing to the CP asymmetry: (a) tree-level, (b) self-energy, and (c) vertex diagrams.}
\label{fig:cpasymmetrydiagrams}
\end{figure}

In leptogenesis, the matter–antimatter asymmetry originates from $CP$-violating decays of heavy right-handed neutrinos. 
At one-loop order, interference between the tree-level and loop amplitudes in the decays $N_i \to \ell_\alpha H$ and $N_i \to \bar{\ell}_\alpha H^\dagger$ generates a net lepton asymmetry. The relevant decay processes are depicted in Figure \ref{fig:cpasymmetrydiagrams}.
This CP violation is quantified by the CP violating parameter $\epsilon$, defined in terms of decay rates:
\begin{equation}
    \epsilon_i = \frac{\Gamma(N_i \rightarrow L H) - \Gamma(N_i \rightarrow \bar{L} H^\dagger)}{\Gamma(N_i \rightarrow L H) + \Gamma(N_i \rightarrow \bar{L} H^\dagger)},
\end{equation}
where $\Gamma$ represents the decay rate of the processes. The precise value of $\epsilon$ is dependent on the masses of the right-handed neutrinos and their Yukawa couplings \cite{Covi_1996, Buchm_ller_2005, Giudice_2004, Di_Bari_2012}
\begin{equation}
    \epsilon_i = \frac{1}{8 \pi}\sum_{j\neq i}\frac{\rm Im[(y^\dag y)_{ij}^2]}{(y^\dag y)_{ii}}  f\left(\frac{M_j^2}{M_i^2}\right)\ .
    \label{eq: epsilon}
\end{equation}
where the function $f(x)$ captures the dependence on the mass ratio of the right-handed neutrinos
\begin{equation}
    f(x) = \sqrt{x} \left[(1 + x) \log\left(\frac{1 + x}{x}\right) - \frac{2 - x}{1 - x}\right].
\end{equation}
From this expression, we observe that $\epsilon$ becomes large when the masses of two right-handed neutrinos are nearly degenerate, in extreme mass degeneracy this leads to the phenomenon called resonant leptogenesis \cite{Pilaftsis_2004}. This regime requires a refined treatment of the self-energy contribution, which dominates in the nearly degenerate limit and develops a regulated enhancement. 
Several prescriptions for regularising this divergent behaviour have been proposed in the literature~\cite{Klaric_2021, Pilaftsis_2004, Riotto_2007, Garbrecht_2014, Garny_2013, Anisimov_2006}. 
Given this theoretical ambiguity, we refrain from analysing the resonant regime in the present study. 
Accordingly, we impose the condition that the mass splitting between the right-handed neutrinos remains much larger than their decay widths~\cite{Moffat_2018}.\\
If instead we take the hierarchical mass limit $M_1\ll M_{2,3}$ then we can maximise the CP asymmetry giving the famous Davidson-Ibarra bound \cite{Davidson:2002qv}.
\begin{equation}
    \epsilon^{\rm max}=\epsilon^{DI}=\frac{3M_1 m^{\rm max}}{16\pi v_H^2}
\end{equation}
where $m^{\rm max}$ is the maximum active neutrino mass, we will assume a hierarchy in the active neutrino mass spectrum meaning we have $m^{\rm max}=0.05\rm eV$. This famous result is independent of the corresponding effective neutrino mass, $\tilde m_1$. This result is derived under the assumption of a hierarchical right-handed neutrino masses; the bound can be dramatically exceeded in the case of nearly degenerate masses. Such scenarios, however, require a degree of fine-tuning and will be discussed later in section \ref{sec: low scale}.

\subsection{The Boltzmann Equations}
To compute the baryon asymmetry, we solve the Boltzmann equations governing the evolution of the lepton asymmetry and the relevant particle abundances. The lepton asymmetry equation contains two distinct contributions: a source term, which generates the asymmetry through $CP$-violating decays, 
and a washout term, which accounts for inverse decays and $\Delta L = 2$ scattering processes \cite{Fong_2012, Ulysses, Ulysses2}. The other bolztmann equations describe the abundances of the lightest right-handed neutrino and the $B-L$ breaking scalar
\begin{equation}
\begin{aligned}
    \frac{dY_\phi}{dz}&= -D_\phi(z)\Big(Y_\phi-Y_\phi^{\rm eq}(z)\Big) \\
    \frac{dY_N}{dz}&=2D_\phi(z)\Big(Y_\phi-Y_\phi^{\rm eq}(z)\Big)-D_N(z)\Big(Y_N-Y_N^{\rm eq}(z)\Big) \\
    \frac{dY_{B-L}}{dz}& = \epsilon D_N(z) \Big(Y_N - Y_N^\text{eq}(z)\Big) - W_N(z) Y_{B-L}\ .
\end{aligned}
\end{equation}
where $z=M_1/T$ is the standard dimensionless variable parametrising the evolution, $D_i$ ($W_i$) denotes the decay (washout) of the asymmetry and are standard results ~\cite{Moffat_2018}  
\begin{equation}
D_i(z) = K_i \, x_i \, z \, \frac{K_1(z_i)}{K_2(z_i)} , \quad W_i(z) = \frac{1}{4} K_i \, \sqrt{x_i} \, K_1(z_i) \, z_i^3
\end{equation}
with $K_1$ and $K_2$ the modified Bessel functions of the first and second kind,  
\begin{equation}
x_i \equiv \frac{M_i^2}{M_1^2}, 
\qquad 
z_i \equiv \sqrt{x_i}\, z,
\end{equation}
and
\begin{equation}
K_i \equiv \frac{\tilde{\Gamma}_i}{H(T=M_i)}\propto\tilde m_i, 
\qquad 
\tilde{\Gamma}_i = \frac{M_i \,(Y^\dagger Y)_{ii}}{8\pi}\ .
\label{eq:Ki}
\end{equation}
Here $H$ is the Hubble expansion rate assuming radiation domination. The $Y_i^{\rm eq}$ terms are the equilibrium abundances of the i-th particle,
\begin{equation}
Y_i^{\mathrm{eq}} = \frac{45}{4\pi^4} \, \frac{g_i}{g_*} \, z_i^2 \, K_2\!\left(z_i\right),
\end{equation}
The final baryon asymmetry is then proportional to $Y_{B-L}$ through the sphaleron conversion factor
\begin{equation}
    Y_B=\frac{28}{79}\ Y_{B-L}\ .
\end{equation}
To parametrise the effect of washout processes, it is convenient to introduce the efficiency factor $\kappa$, defined such that
\begin{equation}
Y_B = \frac{28}{79}\,(Y_N^i + 2Y_\phi^i)\,\epsilon\,\kappa.
\end{equation}
The efficiency factor $\kappa$ quantifies the fraction of the initially produced asymmetry that survives washout effects. Washout strength is controlled by the parameter $K_i \propto \tilde{m}_i$; weak (strong) washout corresponds to $K_1 \ll 1$ ($K_1 \gg 1$). In the absence of washout, $\kappa = 1$, while for strong washout $\kappa \ll 1$. Gauge interactions when present must also be included, namely the processes mediated by the associated $Z'$ gauge boson. However, in Appendix \ref{sec:appendix} we show that for the entire parameter range relevant to our analysis, these gauge-mediated processes are negligibly small. As a result, the leptogenesis dynamics are effectively identical in the global and gauged versions of the theory. The striking consequence is that gravitational waves become the only phenomenological handle capable of distinguishing the two scenarios due to their distinct GW spectral shape, see Fig. \ref{fig:CS benchmark}. We now solve this set of Boltzmann equations analytically to find the lower bound on right-handed neutrino mass for thermal and non-thermal initial abundances.

\subsubsection*{Thermal Leptogenesis Analytical Lower bounds}
\begin{figure}[H]
    \centering
    \includegraphics[width=\textwidth]{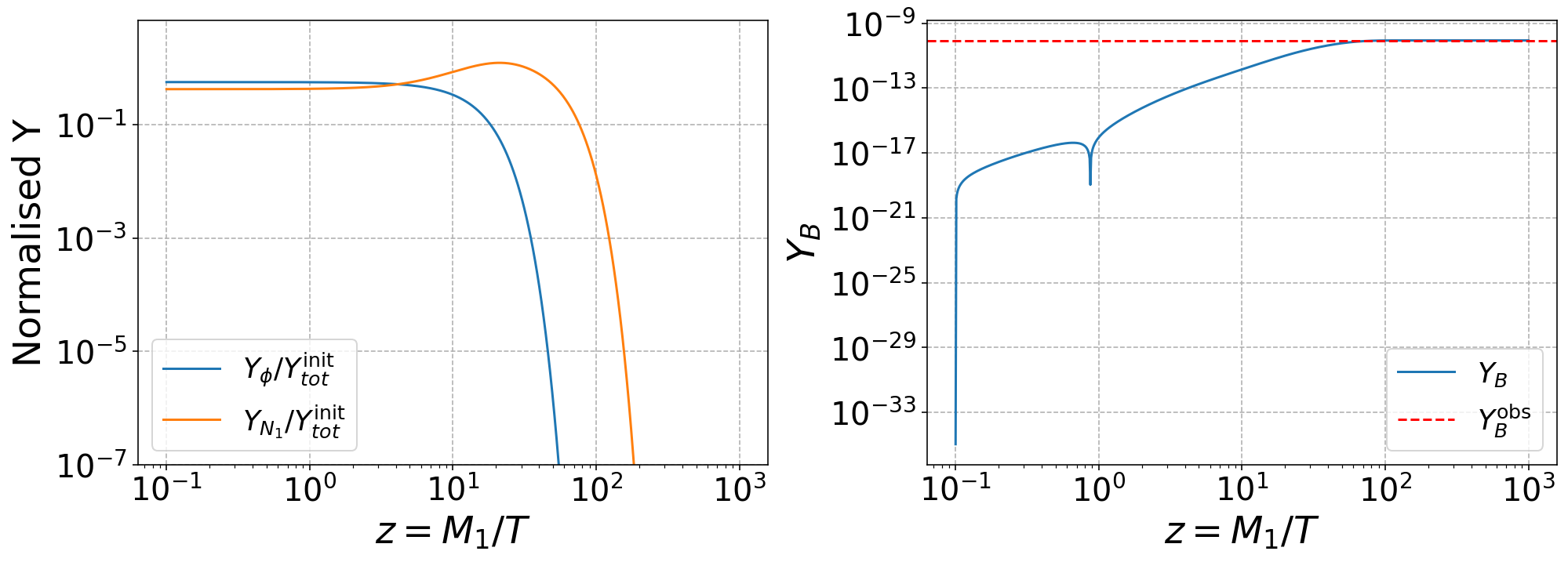}
    \caption{\it  Benchmark result. The parameters are $M_{\phi} = 10^{10}$ GeV, $y_1 = 10^{-5}$ and $M_1=1.74\times 10^8\rm\ GeV$. We take a very small effective neutrino mass $\tilde m_1=10^{-6}\rm\ eV$ to ensure we are in the weak washout regime, the efficiency factor is unity in this benchmark. \textbf{Left:} Normalised abundances of the symmetry breaking scalar $\phi$ and right-handed neutrino. \textbf{Right:} The baryon asymmetry. This shows with a $U(1)_{\rm B-L}$ symmetry thermal leptogenesis can be achieved at mass range $\mathcal{O}(10^8)\rm\ GeV$, nearly an order of magnitude below the conventional Davidson-Ibarra bound.}
    \label{fig:benchmark}
\end{figure}

In this subsection, we aim to identify the parameter space in which leptogenesis is successful which means deriving a lower bound on the right handed neutrino mass. Since the maximal CP asymmetry scales with the right-handed neutrino mass, this imposes a lower bound on that mass. This mass is directly proportional to the vev of $\phi$ which is observable in the the cosmic string background linking the viable parameter space to the detectability. We therefore consider the ideal scenario to minimise the lower mass bound in order to define the parameter space for successful leptogenesis. In this scenario the scalar field has a mass hierarchy $M_1<\frac{1}{2}M_\phi<\frac{1}{2}M_{2,3}$. This means all the thermal initial abundance of the scalar field decay into only the lightest right-handed neutrino. Then we also assume washout is negligible then the  Bolztmann equations collapses down to 
\begin{equation}
    Y_B\leq \frac{28}{79}\ \epsilon^{DI}\ (Y_N^{i}+2Y_\phi^i)
\end{equation}
with,
\begin{equation}
\begin{aligned}
    Y_N^{i}=3.9\times 10^{-3},&\quad Y_\phi^{i}=5.2\times 10^{-3},\\
    \epsilon^{DI}=\frac{3 M_1 m_3}{16\pi v_H^2},\ &\quad Y_B=8.7\times 10^{-11}\ .
\end{aligned}
\end{equation}
where the factor of two comes from there being two right-handed neutrinos produced from one scalar decay. This is easily solved for the lower bound on the right-handed neutrino mass.
\begin{equation}
    M_1 \;>\; 1.74\times 10^{8}\,\text{GeV}
    \label{eq: thermalbound}
\end{equation}
 This is nearly an order of magnitude lower than vanilla leptogenesis due to the extra production of right-handed neutrinos from the decay of $\phi$.
It is also common to consider the case of \emph{zero initial conditions} for the sterile neutrino abundance, $Y_N^i=0$. In this case only the $\phi$ contribution remains, leading to the slightly stronger bound
\begin{equation}
    M_1 \;>\; 2.4\times 10^{8}\,\text{GeV}.
    \label{eq: zerobound}
\end{equation}
To illustrate these results quantitatively, we present a representative benchmark scenario that captures the key features of leptogenesis in this model by numerically solving the bolztmann equations with thermal initial conditions for a set of parameters given in the caption of Fig.~\ref{fig:benchmark}. 
This benchmark demonstrates that the scale of successful leptogenesis can be lowered by roughly an order of magnitude compared to the standard hierarchical case. 

This benchmark serves to emphasise how the presence of a global $U(1)_{B-L}$ symmetry can open new viable regions of parameter space, lowering the required seesaw mass scale. 
It provides a concrete reference point for comparison with more general parameter scans presented later in this work.

\subsubsection*{Non-Thermal Leptogenesis Analytical Lower Bounds}

If instead of thermal production of $\phi$ we consider non-thermal production the scale of leptogenesis can be lowered by several orders of magnitude. Heavy Majorana neutrinos generate radiative corrections to the Higgs mass parameter and to avoid introducing fine-tuning in the Higgs sector requires the lightest Majorana mass to lie below $\mathcal{O}(10^{7\text{--}8}\,\mathrm{GeV})$. Working at lower scales therefore helps keep the electroweak sector natural \cite{Vissani_1998}. Assuming a large non-thermal initial abundance of the scalar field such that $Y_\phi^i \gg Y_N^i$, and once again taking washout to be zero, the baryon asymmetry simplifies to
\begin{equation}
    Y_B \;\simeq\; \frac{56}{79}\,\epsilon^{DI}\,Y_\phi^i,
\end{equation}
which can be inverted to give the lower bound on the right-handed neutrino mass,
\begin{equation}
    M_1 \;>\;
    \frac{16\pi v_H^2\,Y_B}
    {\displaystyle \frac{56}{79}\,3\,m_3\,Y_\phi^i}
    \;\simeq\;
    \frac{1.25\times10^{6}\,\mathrm{GeV}}{Y_\phi^i}.
\end{equation}
Thus, for a large non-thermal abundance, the lower bound scales inversely with $Y_\phi^i$, allowing successful hierarchical leptogenesis down to  $\mathcal{O}(10^{6})~\mathrm{GeV}$ for $Y_\phi^i = \mathcal{O}(1)$. We impose this rough upper limit for two reasons. The first is to prevent the species from ever dominating the energy density of the Universe. The second is the largest non-thermal production mechanism is from inflaton decays, so this serves as a natural benchmark for the maximum allowed initial abundance. This demonstrates that non-thermal production can drastically relax the mass requirement compared to the standard thermal scenario. \\
We also need to reconsider the condition needed to avoid early matter domination by $\phi$ particles. To avoid a $\phi$-dominated era we require that its decay rate of $\phi$ into right-handed neutrinos exceeds the Hubble rate at the onset of domination,
\begin{equation}
    \Gamma_\phi > H(T_{\rm dom}^\phi),
\end{equation}
which gives the condition in terms of the Majorana mass coupling $y_i$
\begin{equation}
    y_i >
    \sqrt{
    8\pi\times1.66\times\sqrt{g_*}
    \left(\tfrac{4}{3}\right)^2
    \frac{M_\phi (Y_\phi^i)^2}{M_{\rm Pl}}
    }
    \;\simeq\;
    7.9\times10^{-9}\,
    Y_\phi^i\,\sqrt{\frac{M_\phi}{\mathrm{GeV}}}.
    \label{eq:y_avoid_MD}
\end{equation}
For a large non-thermal abundance $Y_\phi^i = \mathcal{O}(1)$, this becomes
\begin{equation}
    y_i \gtrsim 7.9\times10^{-9}\,\sqrt{\frac{M_\phi}{\mathrm{GeV}}}.
\end{equation}
Thus, a sufficiently large Yukawa coupling is required to prevent $\phi$ from dominating the energy density of the Universe prior to its decay. We note that a $\phi$ dominated era is interesting in more than one respect, as well as altering the dynamics of leptogenesis it will also modify the gravitational wave background, by introducing characteristic spectral features \cite{Datta:2025vyu}. Specifically the mass and decay rate of $\phi$ could be determined by the gravitational wave background. These signatures offer a potential observational window into the early Universe and could directly link the cosmological gravitational-wave spectrum to the parameters governing leptogenesis. A detailed investigation of this connection is currently in progress and will be presented in forthcoming work.

\subsection{What about Flavour Effects?}

\begin{figure}[H]
    \centering
    \includegraphics[width=0.75\textwidth]{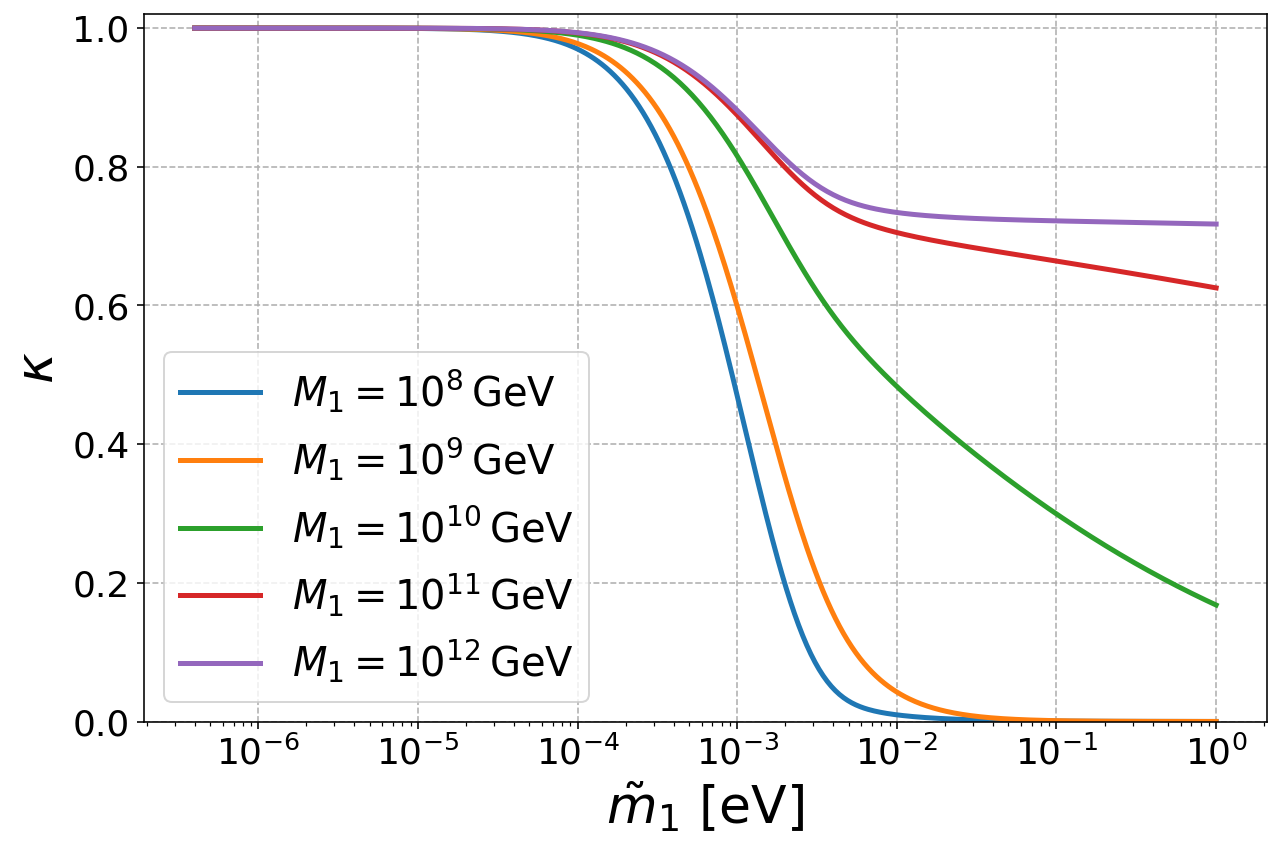}
    \caption{\it Efficiency factor $\kappa$ as a function of the effective neutrino mass $\tilde m_1$ for different values of the right-handed neutrino mass $M_1$. The fixed parameters are $M_{\phi} = 2\times 10^{12}\,\text{GeV}$ and $y = 10^{-4}$. For small $\tilde m_1$, the efficiency saturates to unity, while at larger values washout effects suppress $\kappa$ more strongly, with the onset of suppression depending on $M_1$. Deviation from the weak washout regime, and therefore the point at which we need to consider flavour effevcts occurs around $\tilde m_1 \approx 10^{-4.5}\,\text{GeV}$. This means our analytical bounds are valid for $\tilde m_1\lesssim 10^{-4.5}\rm\ eV$.}
    \label{fig:flavourless_mtilde_scan}
\end{figure}

Flavour effects \cite{Nardi_2006, Nardi:2006fx, Abada_2006, Antusch_2006, Blanchet_2007, DeSimone:2006nrs, Cirigliano_2010, Simone_2007, Racker_2012, Moffat_2018, baker2024hotleptogenesis, Ulysses, Ulysses2, blanchet2013leptogenesisheavyneutrinoflavours} play a crucial role in a complete treatment of leptogenesis, as they determine how individual lepton flavours evolve and how washout processes act differently in each sector. In regimes of strong washout, these differences can significantly alter the final lepton asymmetry and lower the scale of leptogenesis, making a flavoured treatment essential. Conversely, when the overall washout is weak, flavour effects can be safely neglected, and the unflavoured Boltzmann equation provides an accurate description of the dynamics.  To identify the onset of significant washout, in figure \ref{fig:flavourless_mtilde_scan} we plot the leptogenesis efficiency factor, $\kappa$, as a function of the effective neutrino mass, $\tilde{m}_1$, by numerically solving the Boltzmann equations in Eq.~(34) for each value of $\tilde{m}_1$, for fixed benchmark choices of the remaining parameters and for various right-handed neutrino masses. A noticeable deviation of $\kappa$ from unity signals the transition to the regime where washout and therefore flavour effects become important and must be included in the analysis.

Therefore, if we consider the parameter space in the range 
$10^{-4.5}\,\mathrm{eV} \lesssim \tilde{m}_1$, 
we must move into the flavoured regime, which will be explored in the next section. 
An interesting feature of this framework is that the efficiency also depends on $M_1$. 
This dependence is characteristic of extended leptogenesis models that include a non-thermal production mechanism for $N$. 
For smaller sterile neutrino masses, and for fixed values of $M_\phi$ and $y$, the decays of $\phi$ into $N$ occur at smaller values of $z$. 
As a result, a larger fraction of the lepton asymmetry produced during this period is washed out compared to the case where these decays occur later. 
Consequently, for smaller $M_1$ the efficiency is reduced when the other parameters are held fixed. We therefore conclude that for 
$\tilde{m}_1 < 10^{-4.5}\,\mathrm{eV}$, 
successful thermal leptogenesis can be achieved for 
$M_1 > 1.74 \times 10^{8}\,\mathrm{GeV}$ whereas for non-thermal production the mass limit can be reduced down to $M_1>\mathcal{O}(10^{6})\rm\ GeV$.

\medskip

\section{Numerical Results}
\label{sec: numeric leptogenesis}
We now include washout and flavour effects in our analysis by solving the flavoured Boltzmann equations numerically. Including flavour effects not only makes the treatment of leptogenesis more accurate, but can also lower the required leptogenesis scale when washout processes are significant. In doing so, we determine the region of parameter space in the $(\tilde{m}_1, M_1)$ plane where successful leptogenesis can occur, and subsequently identify the portions of this region that are testable through gravitational waves originating from cosmic strings.

\subsection{Flavoured Boltzmann Equations}
\begin{figure}[H]
    \centering
    \includegraphics[width=0.75\textwidth]{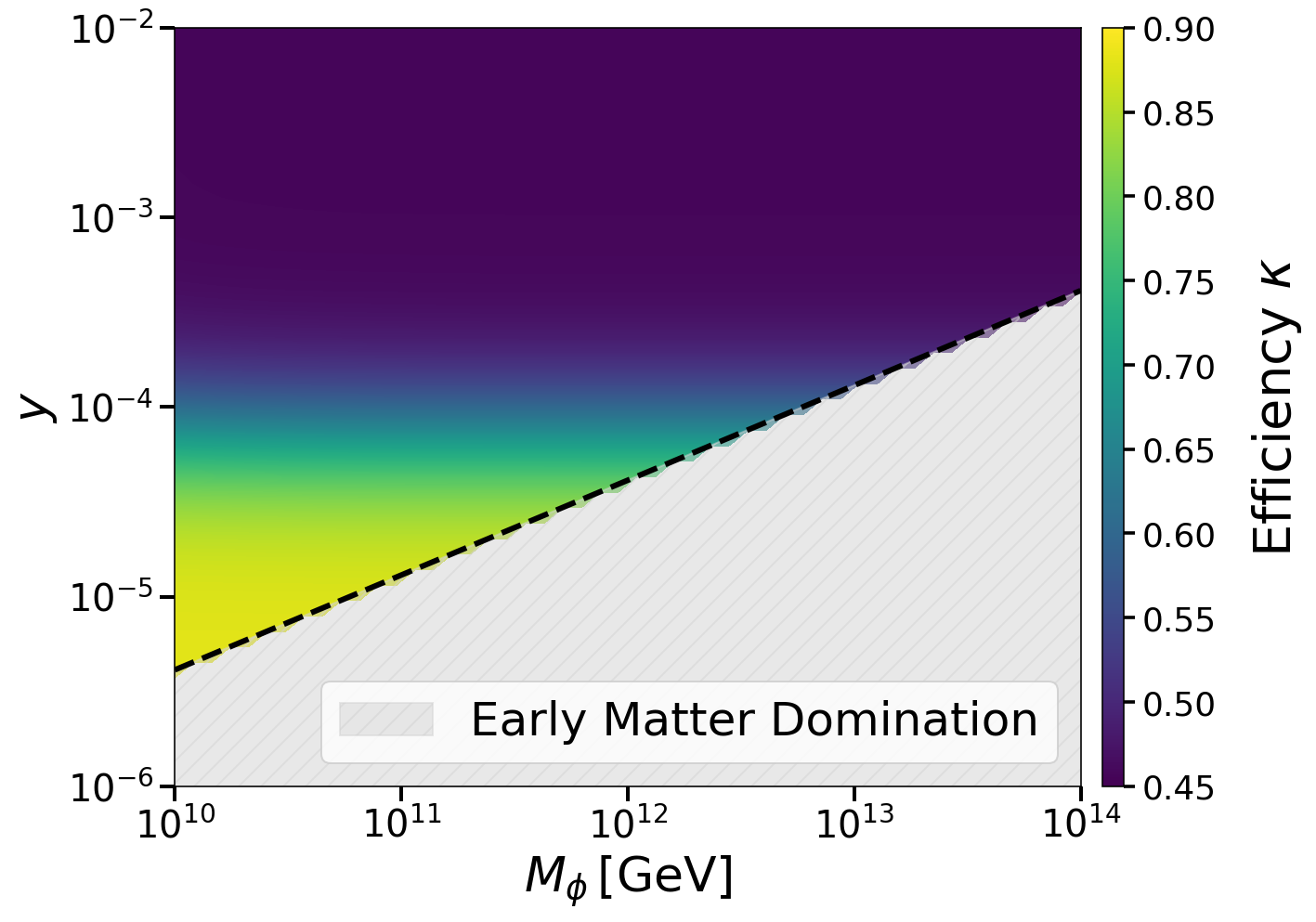}
    \caption{\it Parameter scan to find the optimum values of $M_\phi$ and $y$. The fixed parameters are $M_1 = 10^{9}\,\text{GeV}$ and $\tilde m_{1\alpha} = 10^{-3}$. The values below the bound in Eq. \ref{eq: MD bounds} are excluded as they lead to early matter domination from $\phi$ as discussed in section \ref{sec: rad_dom}. This shows that the optimum yukawa is for the smallest value, which will correspond to also the smallest $M_\phi$ values. This behaviour arises because at larger values of $z$, corresponding to later times, washout processes are weaker; hence, when the decays of $\phi$ occur later (smaller yukawas), fewer of the subsequent $N$ decays are washed out.}
    \label{fig: y and phi values}
\end{figure}

In principle, a fully consistent analysis requires solving the complete set of density matrix equations to account for flavour coherence and oscillations \cite{blanchet2013leptogenesisheavyneutrinoflavours, Moffat_2018, Ulysses, Ulysses2}
However, our present goal is to establish the absolute lower bound on the mass scale required for successful leptogenesis. 
In this regime, corresponding to $M_1 \lesssim 10^9~\mathrm{GeV}$, flavour dynamics are effectively decohered, and it is sufficient to employ the flavoured Boltzmann equations instead of the full density matrix formalism~\cite{Ulysses, Ulysses2}. The equations then take the same form as the flavourless Bolztmann equation only tracking each flavour individually.
\begin{equation}
    \frac{dY_{B-L}^\alpha}{dz} 
    = \epsilon_{1\alpha} D_1(z)\big(Y_{N_1} - Y_{N_1}^{\rm eq}\big)
    - p_{1\alpha} W_1(z) Y_{B-L}^{\alpha},
    \label{eq:flavouredBoltzmann}
\end{equation}
where \(D_1(z)\) and \(W_1(z)\) are the decay and washout terms defined previously for the lightest right-handed neutrino,  and \(p_{1\alpha}\) is the flavour–projection probability which are expressed in terms of the effective neutrino masses
\begin{equation}
    p_{1\alpha} \;=\; \frac{\tilde m_{1\alpha}}{\tilde m_1}, \qquad 
    \tilde m_{1\alpha} \;=\; \frac{|Y_{\alpha 1}|^2 v_h^2}{M_1}, \qquad 
    \tilde m_1 \;=\; \sum_{\alpha=e,\mu,\tau} \tilde m_{1\alpha}\ .
    \label{eq:mtilde_definitions}
\end{equation}
The flavoured CP asymmetry, $\epsilon_{1\alpha}$, is similarly defined in terms of decay rates
\begin{equation}
    \epsilon_{1\alpha}=\frac{\Gamma(N_1\rightarrow L_\alpha H)-\Gamma(N_1\rightarrow \Bar{L_\alpha} H^\dag)}{\Gamma(N_1\rightarrow L_\alpha H)+\Gamma(N_1\rightarrow \Bar{L_\alpha} H^\dag)}\ .
\end{equation}
Using the Casas-Ibarra parametrisation it can be shown that \cite{Abada_2006} each flavour is bounded by the unflavoured Davidson-Ibarra bound and the flavoured effective neutrino masses
\begin{equation}
    \epsilon_{1\alpha}\leq \epsilon_{DI}\ \sqrt{p_{1\alpha}}
\end{equation}
therefore substituting this bound into our flavoured Bolztmann equation and summing over the flavours does not give us back our unflavoured Bolztmann equation and the bounds must be found by a more thorough numerical scan.
The total asymmetry, 
\(\epsilon_1 = \sum_\alpha \epsilon_{1\alpha}\),
is bounded by the usual unflavoured Davidson-Ibarra limit.
To parametrise the flavour structure, we parametrise $(\tilde m_1, \tilde m_{11},\tilde m_{12})$ as the fundamental input variables. Then the third flavour is $\tilde m_{13}=\tilde m_1- \tilde m_{11}-\tilde m_{12}$.  As for the flavoured CP asymmetry parameters $\epsilon$'s we will scan over the allowed values to subject to the condition that they are bound by the sum total and the individual bounds. In the flavoured regime, each lepton flavour $\alpha = e,\mu,\tau$ evolves with its own
CP asymmetry $\epsilon_{1\alpha}$ and washout strength $\tilde m_{1\alpha}$ and therefore has its own efficiency factor $\kappa(\tilde m_{1\alpha})$. The final baryon asymmetry is obtained by summing over all flavours,
\begin{equation}
    Y_{B-L}^{\text{final}} 
    = (Y_N^i+2\ Y_\phi^i)\sum_{\alpha} \kappa_\alpha(\tilde m_{1\alpha})\,\epsilon_{1\alpha}(M_1,\tilde m_1, \tilde m_{11}, \tilde m_{12}),
    \label{eq:YBLequation}
\end{equation}
where $\kappa_\alpha$ is the efficiency factor that quantifies how much of the initially 
generated asymmetry in each flavour survives the corresponding washout. The efficiency depends only on the washout strength and the dynamics of decays and inverse decays, and is therefore independent of the CP asymmetry $\epsilon_{1\alpha}$.  In the unflavoured approximation, the total CP asymmetry satisfies the Davidson-Ibarra 
bound $|\epsilon_1|\leq \epsilon^{DI}(M_1)$ and the individual CP asymmetries obey the constraints
\begin{equation}
    |\epsilon_{1\alpha}| \leq \epsilon^{DI}(M_1)\sqrt{p_{1\alpha}},
    \qquad 
    \abs{\sum_{\alpha} \epsilon_{1\alpha}} \leq \epsilon^{DI}(M_1),
    \label{eq:epsbounds}
\end{equation}
while each flavour experiences a different washout.
Although the sum of CP asymmetries is bounded by the unflavoured limit, the efficiencies are highly non-linear functions of the washout parameters. Consequently, the total asymmetry can exceed the unflavoured value, since the CP asymmetry can be concentrated in the flavour experiencing the weakest washout. This redistribution of CP violation among flavours represents a genuine physical enhancement that is absent in the unflavoured treatment.\\
To investigate this enhancement, we will first determine the flavoured efficiency factors $\kappa_\alpha(\tilde m_{1\alpha})$ by solving the corresponding  Boltzmann equations numerically for each flavour. Once the fitted efficiencies are obtained, the maximal total asymmetry is found by solving the optimisation problem subject to the bounds in Eq.~\eqref{eq:epsbounds}.
By scanning over the flavour fractions $p_{1\alpha}$ and the corresponding CP-asymmetry configurations, we identify the 
optimal distribution that maximises $Y_{B-L}^{\text{flavoured}}$.
Finally, we determine the minimum value of $M_1$ that reproduces the observed baryon asymmetry for each fixed $\tilde m_1$, thereby establishing the lower mass bound in the flavoured regime.\\
We first need to find what which values of $M_{\phi}$ and $y$ give the most efficient baryon asymmetry. We do this by taking a benchmark value of the right-handed neutrino mass and effective neutrino mass and scanning over the Yukawa and mass for the symmetry breaking scalar $\phi$ subject to the condition we avoid early matter domination caused by the the symmetry breaking scalar which we show in Fig. \ref{fig: y and phi values}. 
Figure \ref{fig: y and phi values} demonstrates the smallest values of $y$ and $M_\phi$ will give the largest asymmetry. This behaviour can be understood physically as follows. At larger values of $z$, corresponding to later times in the evolution, the washout processes are less efficient. Therefore, when the decays of the scalar field $\phi$ into right-handed neutrinos $N$ occur later, fewer of the subsequent $N$ decays are subject to washout. We therefore set $M_{\phi}=5M_1$ and $y=4.11\times 10^{-9} \sqrt{M_\phi/GeV}$, the minimum possible Yukawa allowed. This means the scalar decays occur very late, ensuing right-handed neutrino decays take place in a regime with minimum washout. As a result, the lepton asymmetry can be dramatically enhanced, even in parameter regions that would correspond to strong washout in minimal thermal leptogenesis. \\
This mechanism shares a resemblance to leptogenesis during reheating, where right-handed neutrinos are produced from inflaton decay rather than a thermal initial abundance \cite{Ghoshal:2022fud}. In such scenarios the temperature of the universe can always remain well below the sterile neutrino mass, leading to negligible washout. If during reheating the temperature does not remain below the right-handed neutrino mass, however, then washout processes occur. However, in our scenario the Universe can reach temperatures $T \gtrsim M_{1}$ where washout would otherwise be strong, but only a small right-handed neutrino population is present at that stage. Most $N$ are instead produced later via delayed $\phi \rightarrow NN$ decays, so the dominant contribution to the asymmetry arises when $T \ll M_{1}$. In reheating leptogenesis, on the other hand, the temperature has to remain low for the entire relevant period to prevent washout from becoming efficient at any point. 
\subsection{High-scale Thermal Leptogenesis}
\label{HS thermal lepto}
\begin{figure}[H]
    \centering
    \includegraphics[width=0.75\textwidth]{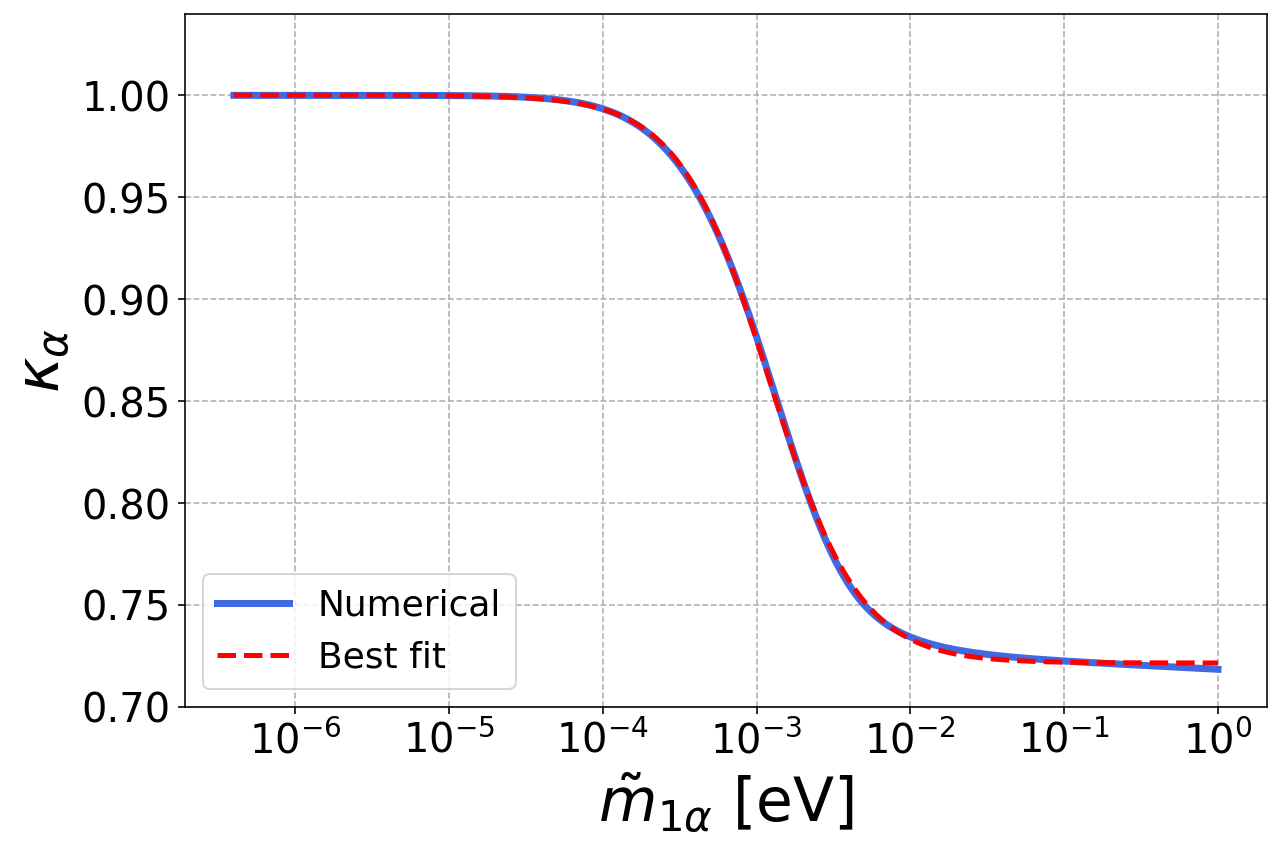}
    \caption{\it Parameter scan showing that, for the optimal values of $M_\phi$ and $y$, the flavour efficiency factor becomes independent of the sterile neutrino mass. As expected the efficiency decreases with the effective neutrino mass parameter corresponding to a stronger washout. However because there are still right-handed neutrinos being produced late the efficiency never goes below $70 \%$. }
    \label{fig:flavoured mtilde scan}
\end{figure}

We now apply the optimal values of the Yukawa coupling and $M_\phi$ to the high-scale leptogenesis scenario. 
With these parameters fixed, the efficiency factor becomes independent of the sterile neutrino mass $M_1$. We numerically scanned over the effective neutrino mass and plot the efficiency and the numerical best fit against flavoured effective neutrino mass in figure \ref{fig:flavoured mtilde scan}.

The efficiency never falls below about $70 \%$ due to the late decaying $\phi$ continuing to inject right–handed neutrinos after washout processes have frozen out. In contrast, in standard thermal leptogenesis, the efficiency at large effective neutrino masses can be suppressed by several orders of magnitude, reaching $\kappa=\mathcal{O}(10^{-4})$. Maintaining $\kappa>0.7$ therefore represents a dramatic improvement in this regime. The results are sufficiently well behaved that it is convenient to give an analytic fit to the numeric results. 
To construct a suitable parameterisation, we note that the efficiency must approach unity for small effective neutrino masses ($\tilde{m}$), while for large $\tilde{m_1}$ it asymptotes to a constant less than unity. We therefore adopt a fit that reproduces this limiting behaviour with a power law interpolation in between,
\begin{equation}
    \kappa(\tilde m_{1\alpha}) = a + \frac{1-a}{1 + b\,\tilde m_{1\alpha}^c} \, .
\end{equation}
The numerical solution for the efficiency factor $\kappa_\alpha(\tilde{m}_{1\alpha})$ is accurately described by the best fit formula
\begin{equation}
    \kappa(\tilde{m_{1\alpha}})
    = 0.721
    + \frac{(1.000 - 0.721)}{1 + 2.083\times10^{4}\,\tilde{m_{1\alpha}}^{1.480}} \,,
    \label{eq:bestfit_thermal}
\end{equation}
which provides an excellent fit to the numerical data with a coefficient of determination $R^{2} = 0.999944$. We now look to solve the optimisation problem discussed. The aim is to maximise the baryon asymmetry,
\begin{equation}
    Y_B=\frac{28}{79}Y^i\sum_\alpha \kappa(\tilde m_{1\alpha})\epsilon(M_1,\tilde m_1, \tilde m_{1\alpha})\label{eq:YB}
\end{equation}
subject to the constraints,
\begin{equation}
    \tilde m_1 =\sum_\alpha \tilde m_{1\alpha}, \quad
    \epsilon_\alpha \leq \epsilon_{DI}\sqrt{\frac{\tilde m_{1\alpha}}{\tilde m_1}}, \quad
    \sum_\alpha\epsilon_\alpha\leq \epsilon_{DI}, \quad \epsilon_{DI}=\frac{3 M_1 m_\nu^{max}}{16\pi v_h^2}\label{eq:constraints}
\end{equation}
We perform a numerical scan over the flavour and CP violating parameters, maximising the baryon asymmetry of \ref{eq:YB} with the constraints \ref{eq:constraints} using an numerically interpolation function constructed from numerically computed values of 
$\kappa$ obtained via repeatedly solving the Boltzmann equations for different values of $\tilde{m}_{1 \alpha}$, and subsequently scan over $M_1$ to obtain the maximum baryon asymmetry for a fixed right-handed neutrino mass and effective neutrino mass. We show the results of this scan in figure \ref{fig: high scale scan}.
\begin{figure}[H]
    \centering
    \includegraphics[width=0.75\textwidth]{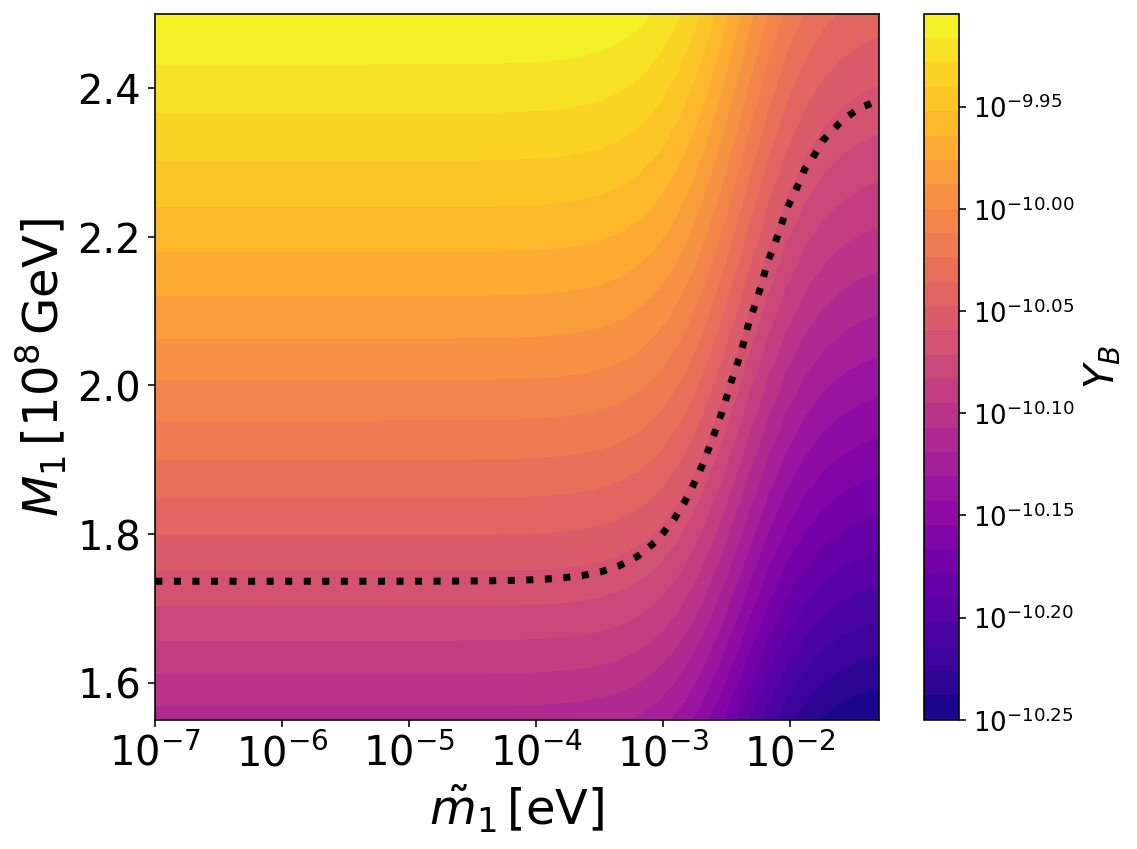}
    \caption{\it Complete parameter scan for successful leptogenesis over $M_1$, $\tilde m_1$. The parameters, $M_{\phi}$ and $y$ have been optimised. For small $\tilde m_1$, the bound for $M_1$ tends to our analytical value for weak washout whereas for the stronger washout regime the lower bound for the mass increases up to the bound for zero initial sterile neutrinos. This shows in our model leptogenesis can be achieve at mass range $\mathcal{O}(10^8)\rm GeV$ independent of washout regime lowering the Davidson-Ibarra mass bound dramatically for large effective neutrino masses.}
    \label{fig: high scale scan}
\end{figure}
We can wee the minimum mass interpolates between the bound for thermal initial conditions for right-handed neutrinos and for zero initial conditions for right-handed neutrinos equations \ref{eq: thermalbound} and \ref{eq: zerobound} respectively. The numerical results are excellently reproduced by the best fit equation
\begin{equation}
    M_1^{\min} \;=\;
    2.40\times10^{8}
    \;-\;
    \frac{7\times10^{7}}
         {1 + 3.38\times10^{3}\,\tilde m_1^{1.50}} \,,
    \label{eq:M1min_fit}
\end{equation}
This demonstrates that, within our model, successful leptogenesis can be achieved at a mass scale of $\mathcal{O}(10^8)\,\mathrm{GeV}$, independent of the washout regime, thereby lowering the Davidson–Ibarra bound by roughly an order of magnitude for small effective neutrino masses but by around 3-4 orders of magnitude for large effective neutrino masses.

\subsection{High-scale Non-Thermal Leptogenesis}
We now repeat our analysis for non-thermal production of $\phi$ with various initial conditions of $Y_\phi^i$. We must also change our yukawa so that there is no early matter domination, the value is taken to be
\begin{equation}
    y = 9\times10^{-9}\,Y_\phi^i\, \sqrt{\frac{M_\phi}{\mathrm{GeV}}}\ .
\end{equation}
We repeat the calculation for figure \ref{fig:flavoured mtilde scan} for various non-thermal initial conditions and show the results in figure \ref{fig: high scale scan}. 
\begin{figure}[H]
    \centering
    \includegraphics[width=0.75\textwidth]{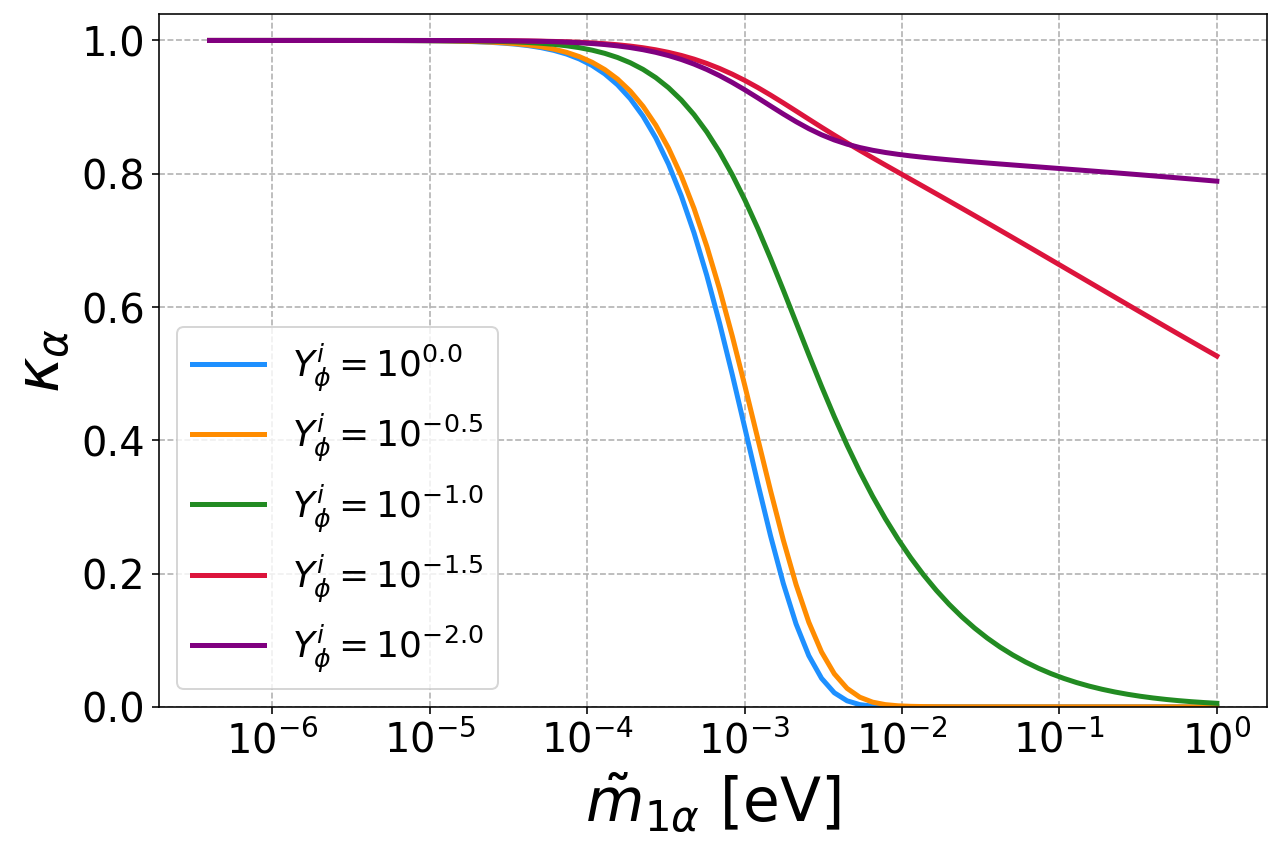}
    \caption{\it Efficiency of leptogenesis as a function of the effective neutrino mass for different non-thermal initial abundances of $\phi$. Smaller initial values of $Y_\phi$ delay the decay of $\phi$ while still avoiding early matter domination, causing right-handed neutrinos to be produced later, after washout has weakened. As a result, the surviving asymmetry is larger and the efficiency increases.}
    \label{fig: high scale scan nontherm}
\end{figure}

As the initial abundance $Y_\phi^i$ decreases, the efficiency factor $\kappa_\alpha$ generally increases. 
This behaviour arises because smaller $Y_\phi^i$ corresponds to a smaller Yukawa coupling and hence to later $\phi$ decays. 
The right-handed neutrinos $N_1$ produced from these decays are then injected into the plasma at a time when washout processes are already less effective, allowing a larger fraction of the generated lepton asymmetry to survive. 
Conversely, for large $Y_\phi^i$, $\phi$ decays occur earlier, when the Universe is still close to thermal equilibrium and washout is strong, leading to a reduced efficiency.

An interesting feature appears around the point $Y_\phi^i =\mathcal{O}(Y_{N}^i)$, where the efficiency curves cross. 
At this transition, the initial thermal population of right-handed neutrinos and the non-thermal contribution from $\phi$ decays become comparable, and the two production channels interfere dynamically. 
For $Y_\phi^i > Y_{N}^i$, the asymmetry is dominated by the early, thermal-like contribution and is more strongly washed out. 
For $Y_\phi^i < Y_{N}^i$, the non-thermal channel becomes dominant, and the later injection of $N_1$ leads to more efficient asymmetry generation. 
This crossover therefore marks the point where the dynamics shift from a washout-dominated (quasi-thermal) regime to a non-thermal, decay-driven regime with enhanced efficiency.\\
The rational–floor model 
\begin{equation}
    \kappa(\tilde{m}) = a + \frac{1 - a}{1 + b\,\tilde{m}^{\,c}}
\end{equation}
provides an excellent fit across all initial abundances $Y_\phi^i$, including the thermal case. 
The fitted parameters $(a,b,c)$ and coefficients of determination $R^2$ are summarised in Table~\ref{tab:kappa_fits}, showing that the model reproduces the numerical results with very high accuracy.

\begin{table}[H]
\centering
\renewcommand{\arraystretch}{1.3}
\setlength{\tabcolsep}{10pt}
\begin{tabular}{|c|c|c|c|c|}
\hline
$Y_\phi^i$ & $a$ & $b$ & $c$ & $R^2$ \\
\hline
\textbf{Thermal} & $0.721$ & $2.083\times10^{4}$ & $1.480$ & $0.999944$ \\
$1.0$       & $0.000$ & $4.421\times10^{5}$ & $1.820$ & $0.999093$ \\
$3.2\times10^{-1}$ & $0.000$ & $1.966\times10^{5}$ & $1.742$ & $0.999211$ \\
$1.0\times10^{-1}$ & $0.067$ & $9.318\times10^{2}$ & $1.148$ & $0.999581$ \\
$3.2\times10^{-2}$ & $0.686$ & $1.408\times10^{2}$ & $0.931$ & $0.997895$ \\
$1.0\times10^{-2}$ & $0.816$ & $1.061\times10^{4}$ & $1.400$ & $0.999848$ \\
\hline
\end{tabular}
\caption{\it Best-fit parameters for the rational–floor model 
$\kappa(\tilde{m}) = a + (1 - a)/(1 + b\,\tilde{m}^{\,c})$
for different initial abundances $Y_\phi^i$, including the thermal case. 
All fits yield $R^2 > 0.997$, confirming the excellent accuracy of the parametrisation. }
\label{tab:kappa_fits}
\end{table}

Now we repeat the analysis for last section completing the optimisation problem numerically with general $Y_\phi^i$, the minimum $M_1$ against $\tilde m_1$ is shown below,
\begin{figure}[H]
    \centering
    \includegraphics[width=0.75\textwidth]{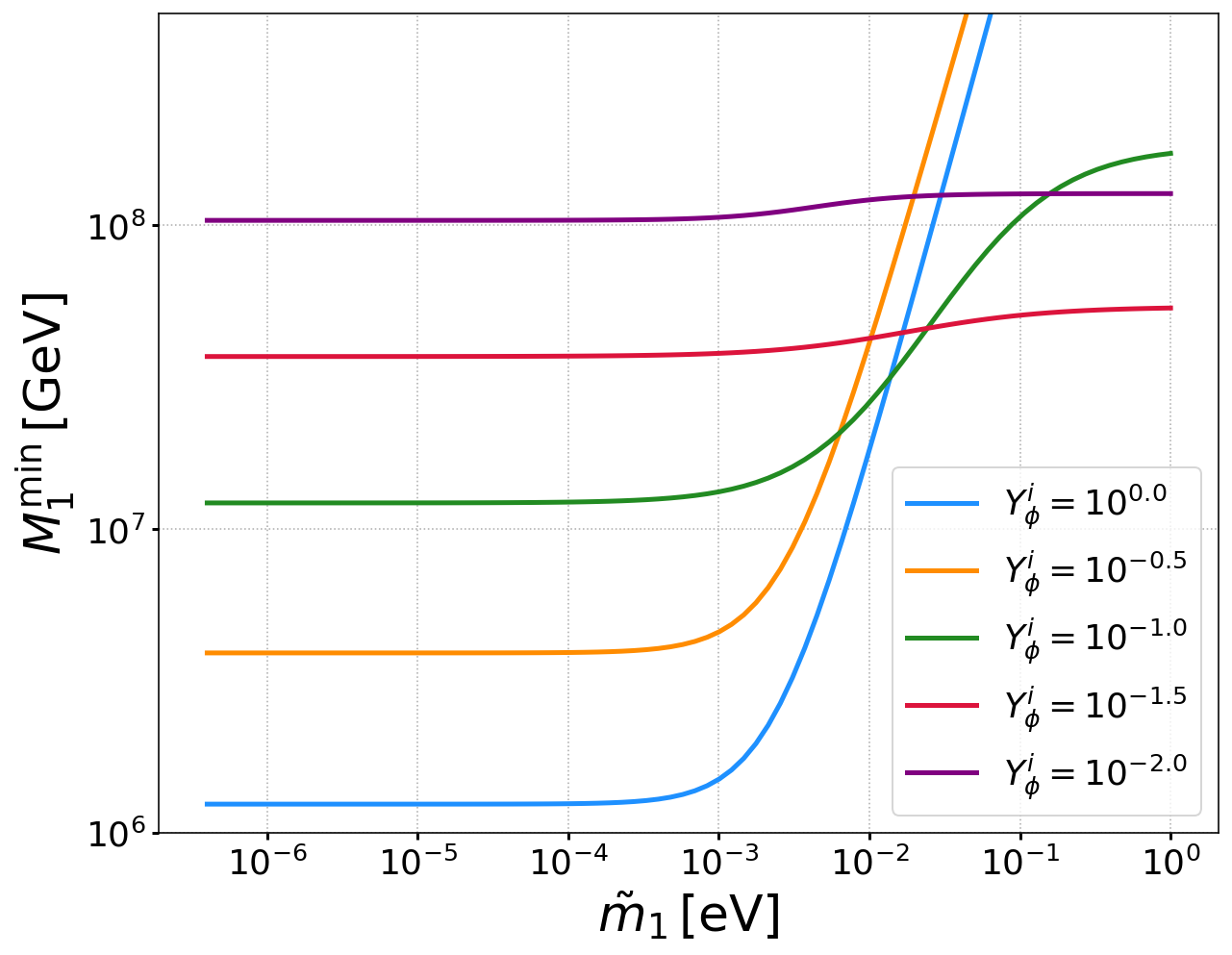}
    \caption{\it Minimum $M_1$ against $\tilde m_1$ for various initial abundances of $\phi$. For larger values of $\phi$ it for negligible washout the values can be reduced inversely proportional to $Y_\phi^i$. however, at larger values the bound on radiation domination means the decays have to occur early and the washout is great so suffers. }
    \label{fig: high scale scan}
\end{figure}
We find that large initial abundances of $\phi$ can substantially lower the mass bound in the weak-washout regime. For small effective neutrino masses, the bound can be reduced
to $\mathcal{O}(10^{6})\,\mathrm{GeV}$ when $Y_\phi$ is very large. At larger values of the effective neutrino mass, however, the constraint from avoiding early matter domination becomes more restrictive: $\phi$ must decay earlier, leaving more time for washout and thus requiring a larger CP-violating parameter to generate the observed asymmetry. As a consequence, the mass bound in this regime is considerably higher.

\subsection{Low-Scale Leptogenesis}
\label{sec: low scale}
In the quasi-degenerate regime where $M_i \simeq M_j$, the loop function $f(x)$ strongly enhances the CP asymmetry. In the limit of extreme mass degeneracy, the self-energy contribution to the CP
asymmetry becomes resonantly enhanced, giving rise to the phenomenon of resonant leptogenesis~\cite{Pilaftsis_2004}. In this regime the standard perturbative treatment breaks down and the self-energy term requires a careful regularisation. Several
prescriptions for regulating this enhancement exist in the literature \cite{Klaric_2021, Pilaftsis_2004, Riotto_2007, Garbrecht_2014, Garny_2013, Anisimov_2006}, but they do not yield a unique, universally accepted expression. Because of this theoretical ambiguity, we do not investigate the resonant regime in the present work. To avoid this region we will enforce that the mass splitting must be greater than one hundred times the decay rate of the right-handed neutrinos.\\  
Expanding the form of epsilon in small $\delta M/M$ gives
\begin{equation}
\epsilon_i \;\simeq\; \frac{1}{16\pi}\,\frac{M_i}{\delta M}\,
\frac{Im\big[(y^\dagger y)_{ij}^2\big]}{(y^\dagger y)_{ii}},\ \leq \frac{1}{16\pi}\,\frac{M_i}{\delta M}\,(y^\dagger y)_{jj}.
\end{equation}
where in the last step we have bounded the numerator with the Cauchy-Schwartz inequality. To remain in the non-resonant regime, the mass splitting must greatly exceed both decay widths \cite{Pilaftsis_2004, Moffat_2018, Spalding:2026jia},
\begin{equation}
\delta M > 100\,\Gamma_1,
\quad
\delta M > 100\,\Gamma_2, \quad \Gamma_i = \frac{(y^\dagger y)_{ii}}{8\pi}\,M_i\ .
\end{equation}
Evaluating $\epsilon_1$, the condition from $\Gamma_1$ yields an upper limit on the CP-violating parameter, expressed in terms of the ratio of effective neutrino masses,
\begin{equation}
|\epsilon_1| \;\lesssim\; \frac{1}{200}\,\frac{(y^\dagger y)_{22}}{(y^\dagger y)_{11}}\,\frac{M_2}{M_1}\simeq\frac{1}{200}\frac{\tilde m_2}{\tilde m_1}\ .
\end{equation}
In contrast, the constraint from $\Gamma_2$ removes any dependence on masses or Yukawas,
\begin{equation}
|\epsilon_1| \;\lesssim\; \frac{1}{200}.
\end{equation}
By symmetry, the corresponding limits for decays of the heavier right-handed neutrino are
\begin{equation}
|\epsilon_2| \;\lesssim\; \min\!\left[\frac{1}{200},
\;\frac{1}{200}\,\frac{\tilde m_1}{\tilde m_2}\right].
\end{equation}
Since both non-resonance conditions must hold simultaneously, the true universal ceiling is
\begin{equation}
|\epsilon_i| \;\lesssim\; \frac{1}{200} 
\end{equation}
Thus, outside the resonant regime, the CP violation parameter cannot exceed the percent level, independently of right-handed neutrino masses and Yukawa couplings. If one chooses to impose the resonance condition $\delta M > b \Gamma_i$ more or less strictly (with $b=100$ our baseline) the corresponding bound on the CP asymmetry follows directly as $\epsilon<1/2b$. These bounds are shown in Table \ref{tab:epsbounds}.

\begin{figure}[H]
    \centering
    \includegraphics[width=\textwidth]{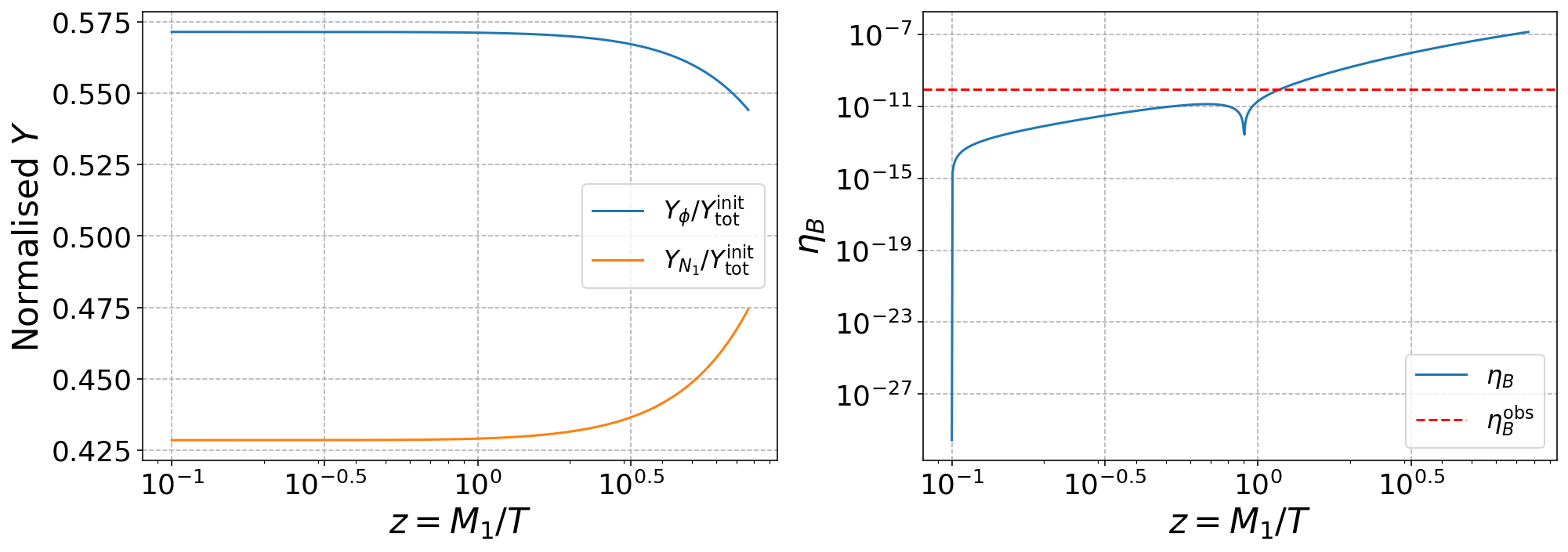}
    \caption{\it Benchmark result for $M_1 = 1~\text{TeV}$ and $\tilde m_1=0.05\rm\ eV$. \textbf{Left:} Normalised abundances of the symmetry breaking scalar $\phi$ and right-handed neutrino. \textbf{Right:} The baryon asymmetry. By the time of electroweak symmetry breaking very few sterile-neutrino decays have occurred, resulting in a low efficiency, $\kappa=0.0055$. Nevertheless, the CP-asymmetry parameter is sufficiently large that the required baryon asymmetry is still readily produced.
    }
    \label{fig: near resonant benchmark}
\end{figure}

\begin{figure}[H]
    \centering
    \includegraphics[width=0.75\textwidth]{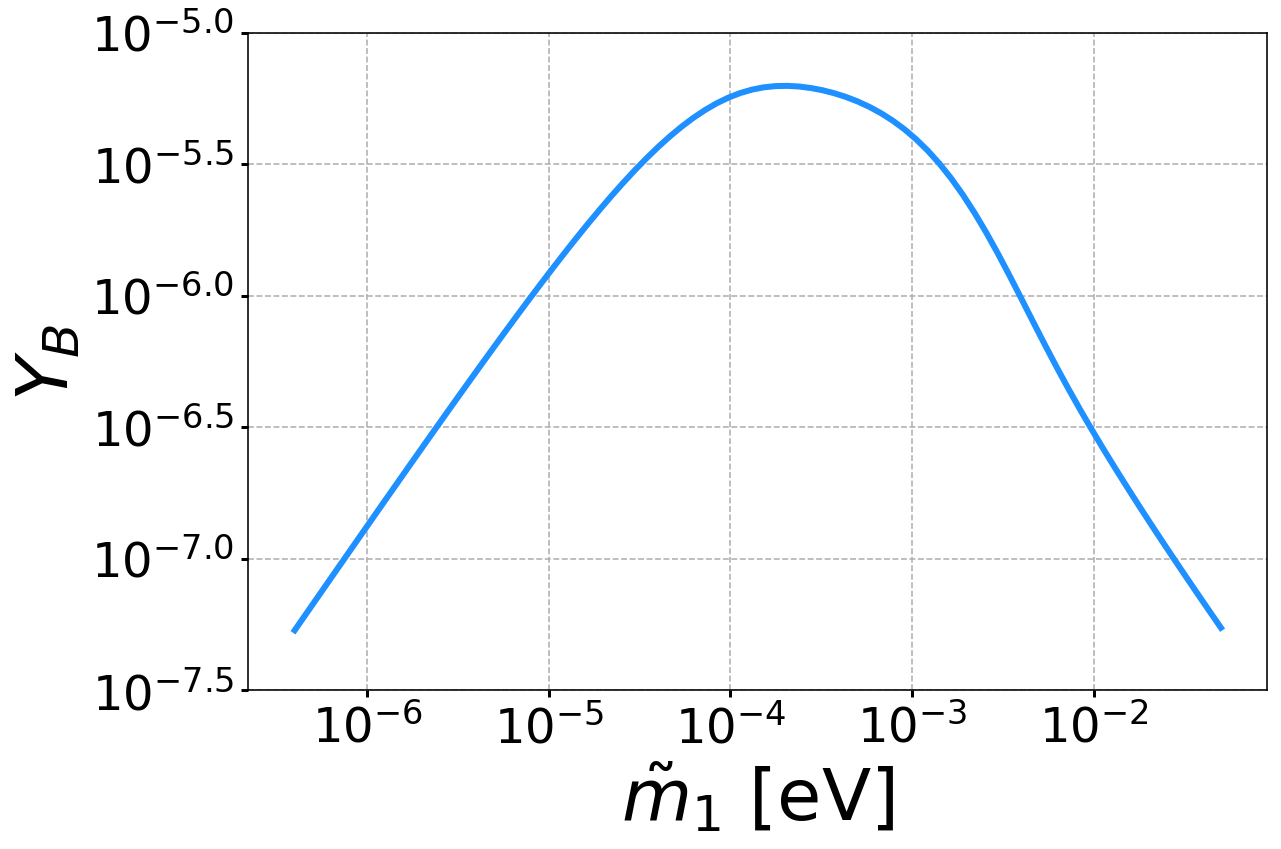}
    \caption{\it Scan over the effective neutrino mass for $M_1 = 1~\text{TeV}$. The mass and Yukawa coupling of the $B\!-\!L$ breaking scalar are chosen to maximise the baryon asymmetry. The plot demonstrates that near-resonant leptogenesis can readily occur for TeV-scale right-handed neutrinos. The suppression at large $\tilde{m}_1$ arises from strong washout effects, while the decrease at small $\tilde{m}_1$ reflects the incomplete decay of right-handed neutrinos before electroweak symmetry breaking. The entire intermediate range of $\tilde{m}_1$, bounded below by the requirement of avoiding early matter domination, yields successful leptogenesis. The observed baryon asymmetry (\ref{eq:BAU}) is below out of range of this plot. 
    }
    \label{fig: near resonant scan}
\end{figure}

\begin{table}[h!]
\centering
\renewcommand{\arraystretch}{1.4}
\setlength{\tabcolsep}{10pt}
\begin{tabular}{|c|c|c|}
\hline
Hierarchy & Bound on $\epsilon_1$ & Bound on $\epsilon_2$ \\
\hline
$\tilde m_1 < \tilde m_2$ 
& $\displaystyle |\epsilon_1| \lesssim \frac{1}{200}$ 
& $\displaystyle |\epsilon_2| \lesssim \frac{1}{200}\,\frac{\tilde m_1}{\tilde m_2}$ \\
\hline
$\tilde m_2 < \tilde m_1$ 
& $\displaystyle |\epsilon_1| \lesssim \frac{1}{200}\,\frac{\tilde m_2}{\tilde m_1}$ 
& $\displaystyle |\epsilon_2| \lesssim \frac{1}{200}$ \\
\hline
\end{tabular}

\captionsetup{justification=raggedright, singlelinecheck=false, font=it}
\caption{Analytic bounds on the CP asymmetries, $\epsilon_{1,2}$, in the quasi-degenerate regime imposing the non-resonant condition $\delta M > 100 \Gamma_i$.}
\label{tab:epsbounds}
\end{table}
We make a simplified analysis just tracking one right-handed neutrino and solving the flavourless bolztmann equations. This is sufficient to show we can go down to $TeV$ scale, as including flavour effects and tracking the decay of a second right-handed neutrino can only increase the lower bound on the mass. With the mass being so low, we cannot assume that all the sterile neutrinos have decayed by the electroweak phase transition so we have no analytical estimates, instead we jump straight into numerics. We first provide a benchmark showing how, even for the strongest washout and the earliest decays, the right-handed neutrino is no where near completing the decays before electroweak symmetry breaking which we fix at $T_{EW}=130\rm\ GeV$. With most of the right-handed neutrino decays occurring after electroweak symmetry breaking the efficiency of leptogenesis is very low, however because the CP asymmetry parameter is so high, successful baryogenesis is still easily achieved. This benchmark is given in Fig. \ref{fig: near resonant benchmark}.
Completing a scan over the effective neutrino mass shows that for any effective neutrino mass in the range we are considering can easily give successful leptogenesis at the TeV scale despite having terrible efficiency factors. This is shown in figure \ref{fig: near resonant scan}.
In our analysis we restrict the scan to $M_1 \gtrsim 10^{3}\,\text{GeV}$. This is because in this low-mass regime the dynamics transition away from standard right-handed neutrino decay leptogenesis and approach the regime where the Akhmedov-Rubakov-Smirnov (ARS) mechanism dominates \cite{Akhmedov_1998, Drewes_2018, Klaric_2021}, that is, where the lepton asymmetry is generated by oscillations of nearly degenerate right-handed neutrinos. Since our focus is on conventional leptogenesis rather than ARS leptogenesis, we conservatively impose the cut-off at $M_1 = 10^{3}\,\text{GeV}$.

\section{Testing Leptogenesis via Gravitational Waves}
In this section we identify the detectable region of the successful leptogenesis parameter space. The amplitude of the gravitational wave spectrum generated by a network of cosmic strings is controlled by the vev $v_{\rm B-L}$, whether the broken symmetry related to the cosmic string is global or local. We therefore express the parameter space in terms of the symmetry-breaking vev, so that the usual mass parameters $(M_1,\tilde m_1)$ are traded for the three independent quantities $(y_{1},\,\tilde m_1,\,v_{B-L})$. We fix the mass of $\phi$ to be an order of magnitude above the right-handed neutrino in order to satisfy $M_2>M_\phi > M_1$ which allows late decays into $M_1$ to be possible while kinematically blocking decays into $M_2$. This allows for the maximum generated asymmetry and true lower bounds for the model as the abundance of $M_1$ is maximized when decays of $\phi$ into $M_2$ are kinematically forbidden.    \par
\begin{figure}[H]
    \centering
    \includegraphics[width=\textwidth]{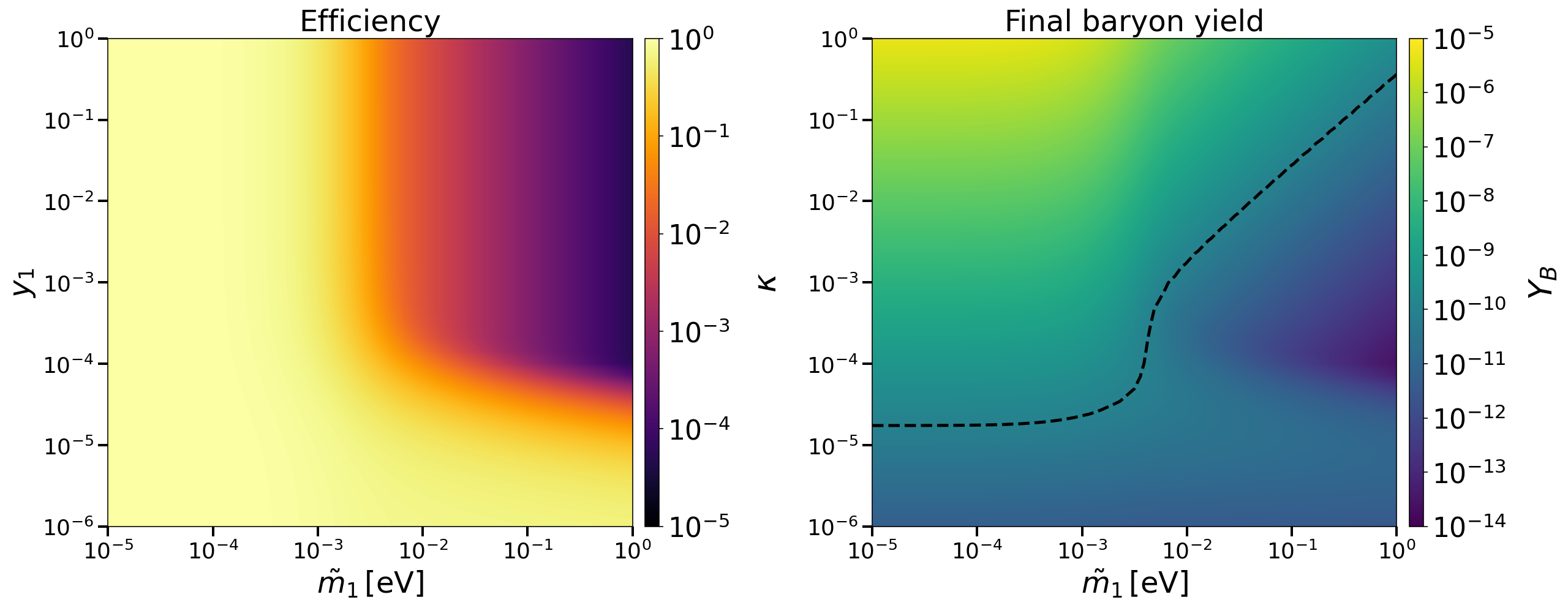}
    \caption{\it Scan over the effective neutrino mass and Yukawa for constant $v_{B-L}=10^{13}\rm\ GeV$. \textbf{Left:} the efficiency factor dependence on Yukawa and effective neutrino mass. \textbf{Right:} The maximum baryon asymmetry taking the Davidson-Ibarra bound for the CP violating parameter and including flavour effects. The efficiency is larger for smaller values of $y_1$ and effective neutrino mass. However the Mass of the right-handed neutrino and therefore the bound on the CP violation parameter grows with the value of the Yukawa coupling in a non-trivial way as the baryon asymmetry is not always monotonic in the Yukawa coupling. The black dashed line denotes the observed baryon asymmetry, any point above this line can have successful baryogenesis.
    }
    \label{fig: vbl benchmark}
\end{figure}

 We will conduct two parameter scans, the first we fix $v_{B-L}$ for various values and vary $y_1$ and $\tilde m_1$, and the second scan we will fix $y_1$ and vary $v_{B-L}$ with the detectable regions for individual detectors. The leptogenesis prediction is independent of whether the symmetry is global or local at the scale of leptogenesis we can probe with cosmic strings. We remind the reader that this issue we give more details about in Appendix \ref{sec:appendix}. As a benchmark example we plot the efficiency and maximum baryon asymmetry in the $y_1, \tilde m_1$ plane for $v_{B-L}=10^{13}$ GeV in Fig. \ref{fig: vbl benchmark}.

\begin{figure}[H]
    \centering
    \includegraphics[width=0.75\textwidth]{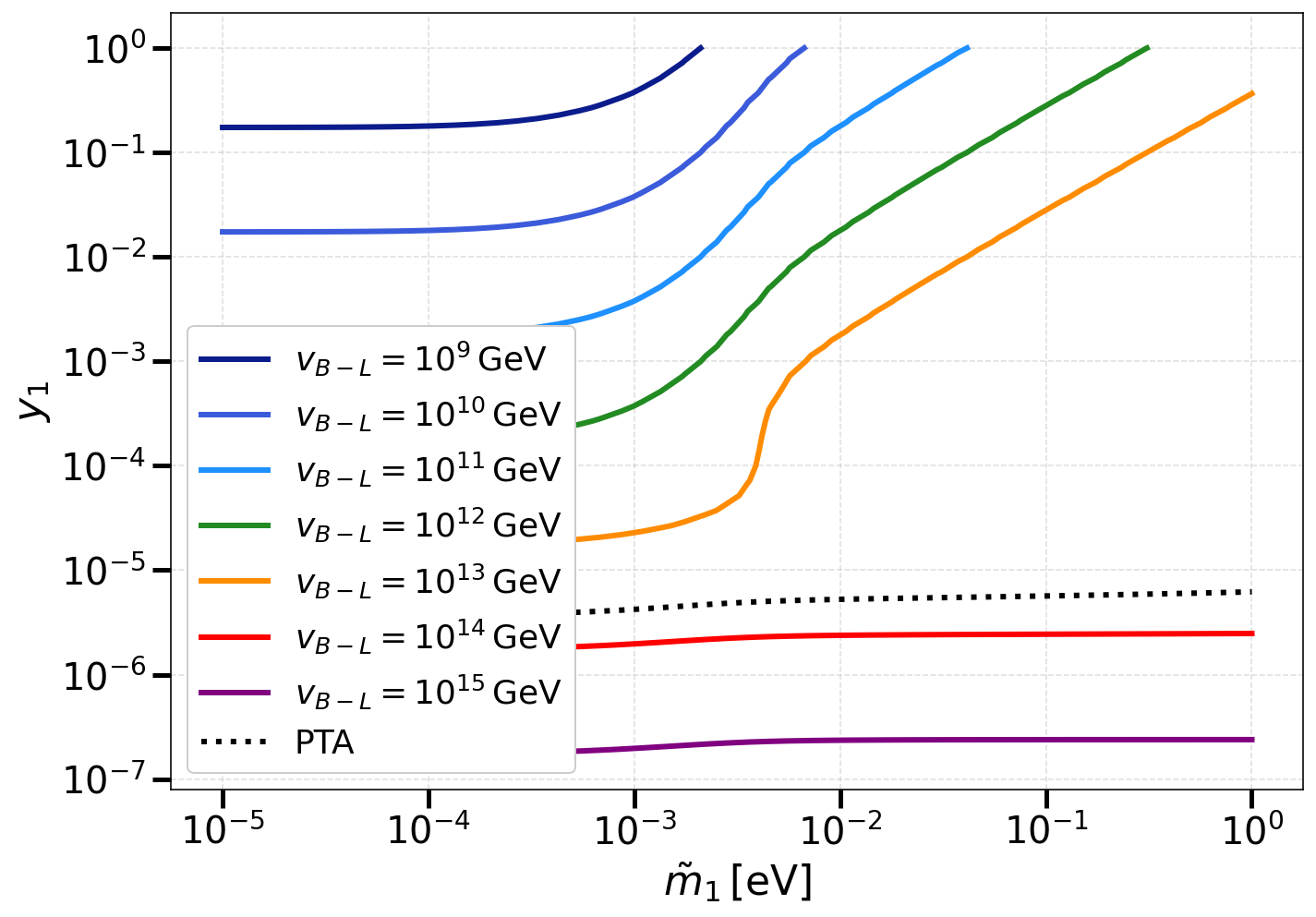}
    \caption{\it Scan over the effective neutrino mass and Yukawa couplings for various values of $v_{B-L}$. Parameter space above the lines give successful leptogenesis including flavour effects for thermal initial abundances. Parameter space above the $v_{B-L}=10^{14},\ 10^{15}$ lines are detectable for the global model and the parameter space above the remaining lines are detectable for the local model. The parameter space below the black dotted line is ruled out for the local model by current gravitational wave measurements. The equivalent bound for global is below this plots range. 
    }
    \label{fig: vbl scan}
\end{figure}

\begin{figure}[H]
\centering
\includegraphics[width=0.9\linewidth]{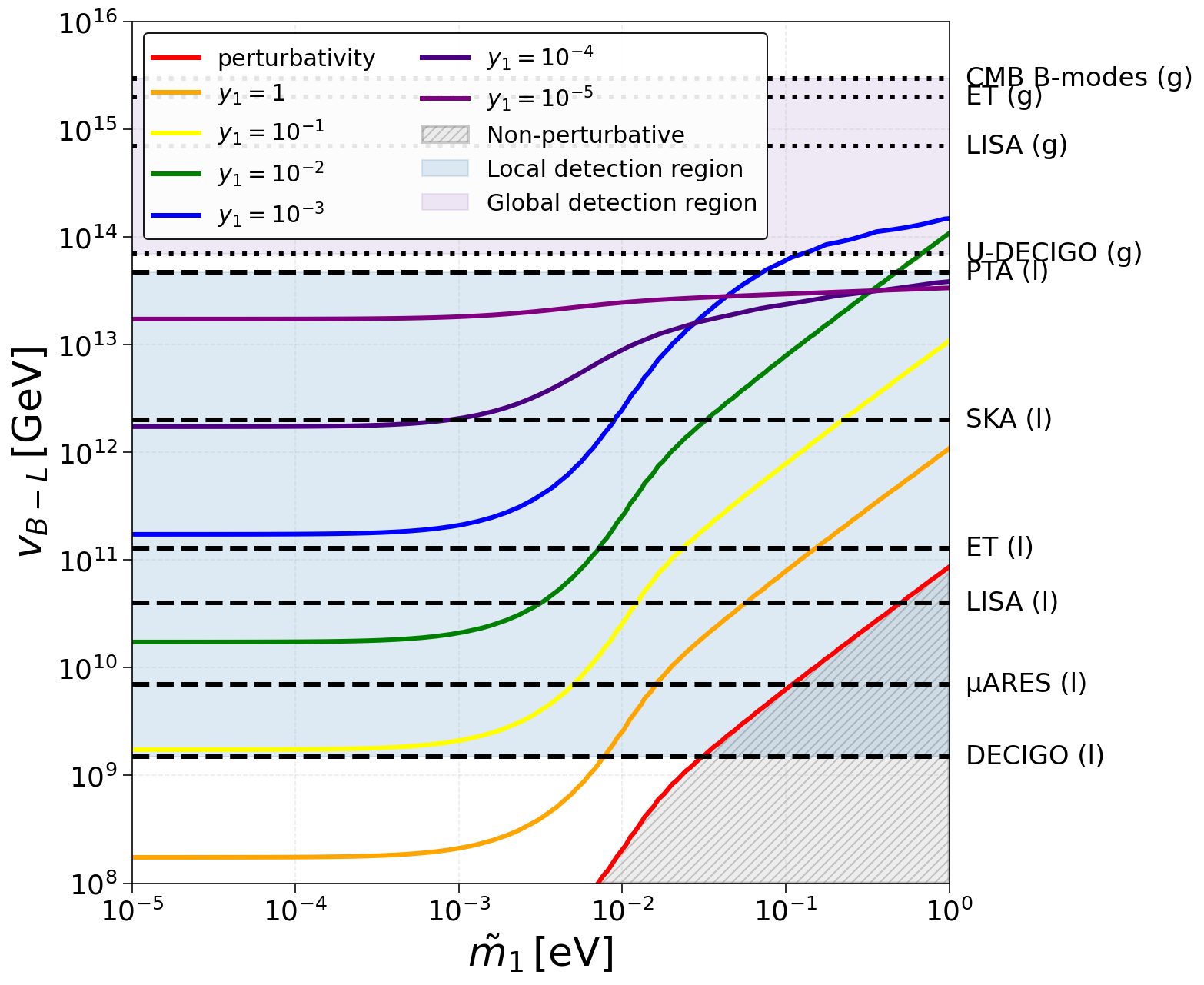}
\caption{\it Contours in the $(\tilde{m}_1,\,v_{B-L})$ plane show where the maximised baryon asymmetry satisfies $Y_B^{\max}=Y_B^{\mathrm{obs}}$ for different values of $y_1$. For each $y_1$, points lying above the corresponding curve generate at least the observed baryon asymmetry. The shaded blue region indicates where local cosmic strings would be detectable, approximately for $1.5\times10^{9}\,\mathrm{GeV} \lesssim v_{B-L} \lesssim 4.7\times10^{13}\,\mathrm{GeV}$. The shaded purple region shows the detectable range for global cosmic strings, roughly $7\times10^{13}\,\mathrm{GeV} \lesssim v_{B-L} \lesssim 3\times10^{15}\,\mathrm{GeV}$. The regions above the CMB B-modes and PTA are the upper bounds for global and local strings receptively with values above this being excluded. We remind the reader that the string tension is directly proportional to $v_{B-L}$. In the dashed region bounded by the perturbativity constraint, successful thermal leptogenesis is not possible.}
    \label{fig:vblscanfull}
\end{figure}

The larger the effective neutrino mass, the lower the efficiency,  as the washout is proportional to $\tilde{m}_1$. This is consistent with existing literature (see e.g. \cite{Buchm_ller_2005} and the references therein). For the RHN-scalar Yukawa coupling $y$ however there is a competition between two effects: the larger the Yukawa the larger the mass of the right-handed neutrino and the higher the bound on the CP violating parameter there is, however it also means that the scalar field decays earlier, making more of the subsequent decays washed out.  We now can repeat this calculation for various $v_{B-L}$ values that produce detectable gravitational wave backgrounds in either global or local $U(1)_{B-L}$ cosmic strings. We plot the $Y_B^{obs}$ line for various $v_{B-L}$ values is shown in figure \ref{fig: vbl scan}.
A complete scan over the effective neutrino mass and the Yukawa coupling allows us to chart the full parameter space in which successful leptogenesis can occur. Larger values of $y$ correspond to larger right-handed neutrino masses, and therefore leptogenesis is naturally driven towards higher values of the $B\!-\!L$ breaking scale. Every region of the cosmic-string parameter space that is potentially detectable by future gravitational-wave observatories contains a corresponding region that realises successful leptogenesis. For roughly $v_{B-L}\in[10^9,10^{13}]~\mathrm{GeV}$, the relevant regions can be probed with local cosmic strings, while for approximately $v_{B-L}\in[10^{14},10^{15}]~\mathrm{GeV}$ the viable parameter space aligns with global strings, the values for individual detectors are shown in figure \ref{fig:vblscanfull}. Taken together, these results demonstrate a robust and testable link between the dynamics of $B\!-\!L$ breaking, the generation of the baryon asymmetry, and the gravitational-wave signatures of cosmic strings. A future detection, or absence of, a stochastic gravitational-wave background from $B\!-\!L$ strings will therefore directly corroborate or exclude large portions of the leptogenesis parameter space, providing one of the most powerful experimental probes of the mechanism studied here.
\\
To show explicitly which gravitational wave detectors can probe which parts of the parameter space we fix the Yukawa and scan over the effective neutrino mass and $v_{B-L}$ and show the maximum baryon asymmetry achievable in figure \ref{fig:vblscanfull}. 
We can see that a larger $y_1$ corresponds to a larger right-handed neutrino mass so the scale of leptogenesis can be lowered to $v_{B-L}=\mathcal{O}(10^8)$ GeV  in congruence with our analytical estimates of the modified Davidson-Ibarra bound, \ref{eq: thermalbound}. For larger Yukawas, the decays of $\phi$ occur earlier so more of the subsequent right-handed neutrino decays are washed out. We now show the lines $Y_B^{\rm max}=Y_B^{\rm obs}$ for various $y_1$ in this plane and also the detectable regions of the parameter space by global or local cosmic strings for various detectors in figure \ref{fig:vblscanfull}, this is a key result as it shows the full lower bounds utilising flavour effects across the majority of the parameter space as well as which regions are detectable or currently excluded. The global cosmic strings lines (for detection) for Cosmic Explorer and DECIGO are not shown as they blur with the line for LISA for global and Einstein telescope for local.  The parameter space above $v_{B-L}=4.7\times 10^{13}$ GeV for local strings are ruled out by PTA measurements, and the parameter space above $v_{B-L}=3\times 10^{15}$ GeV for global strings are ruled out by CMB B-modes.
Very small values of $y_{1}\lesssim 5\times 10^{-5}$ and $y_1\lesssim 10^{-7}$ are already ruled out in the local and global cases respectively. Future gravitational-wave experiments will exclude (or corroborate) further regions of the $(y_{1},\tilde m_1)$ parameter space by testing additional combinations of symmetry-breaking scale and washout strength. A useful aspect of this plot is that, for any chosen $y_{1}$, the intersections between its leptogenesis contour and the detector reach lines immediately give the corresponding values of $M_{1}=y_{1}v_{B-L}$ that the experiment is able to probe. If we enforce $y_1<1$ due to perturbativity and requiring a hierarchy in the right-handed neutrino mass spectrum, for the local theory a non-observation of the GWB from  LISA and Einstein telescope would rule out $\tilde m_1\gtrsim 5\times10^{-2}$ eV and $\tilde m\gtrsim 1.5\times 10^{-1}$ eV respectively. In other words, the plot allows us to read off the effective neutrino mass range associated with each gravitational-wave detector once $y_{1}$ is fixed. Flavour effects become important in this regard as they change the location of the intersection with the detector lines especially in the local scenario moving the intersections to larger $\tilde m_1$ . Flavour effects can rescue parts of the parameter space at large $\tilde m_1$, where the unflavoured treatment would exclude these regions. This provides an additional motivation to include flavour effects: the detectable effective neutrino mass, $\tilde m_1$ window is different once flavour is treated correctly. \\
An interesting feature is that lines cross. This is because the larger the mass of the right-handed neutrino the larger the maximum CP asymmetry however there also the earlier the $\phi$ decays occur, so more of the subsequent decays are washed out. So for a very large effective neutrino masses, $\tilde m_1$, the washout is stronger and can washout more  If we consider non-thermal initial abundances this will just scale the lines vertically, down for larger initial abundances and up for smaller initial abundances. We show the cases for $Y_\phi^i=0.1$ and $Y_\phi^i=0.01$ showing how non-thermal leptogenesis can lower the scale of leptogenesis substantially.
\begin{figure}[H]
\centering
\includegraphics[width=0.49\linewidth]{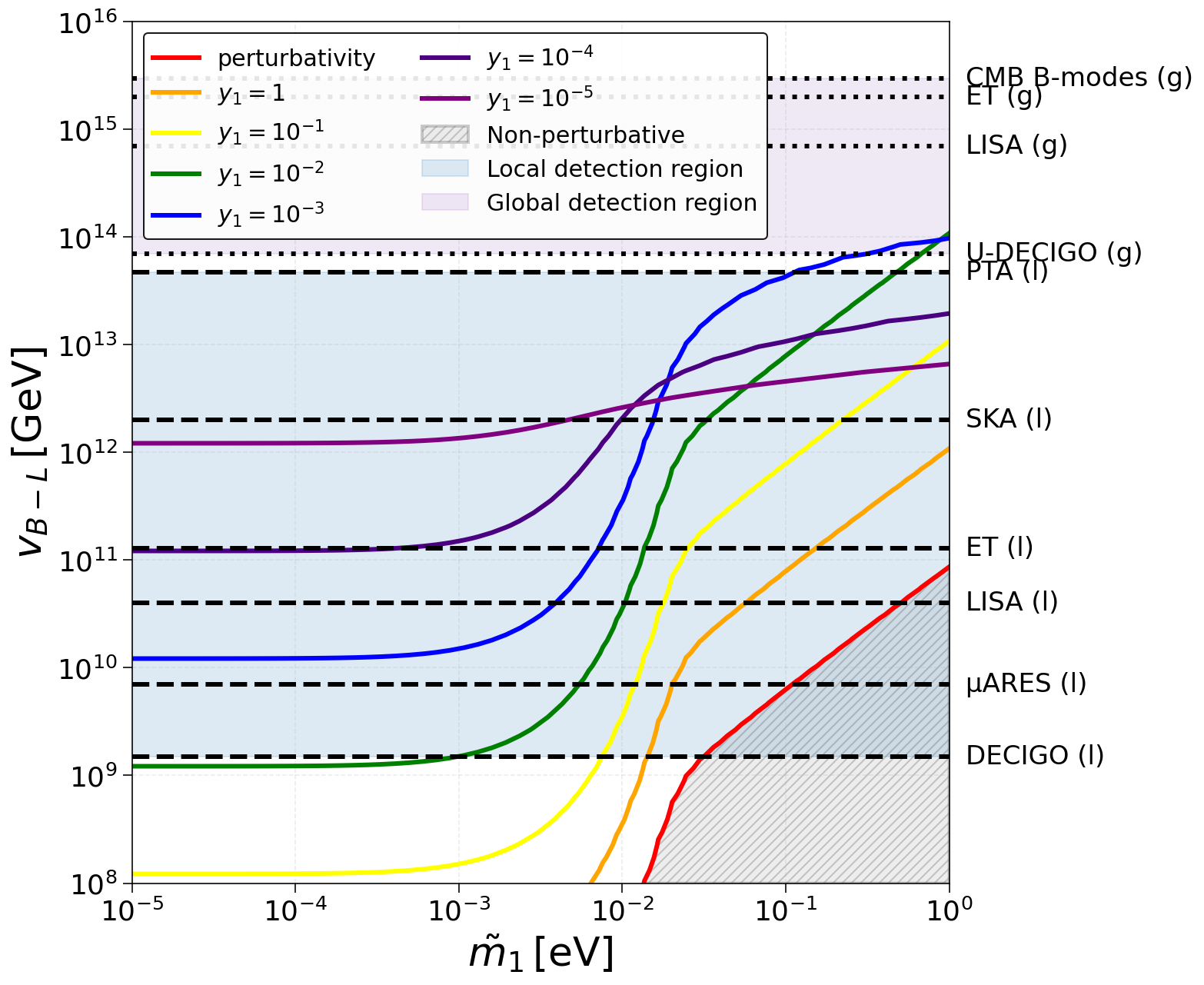}
\includegraphics[width=0.49\linewidth]{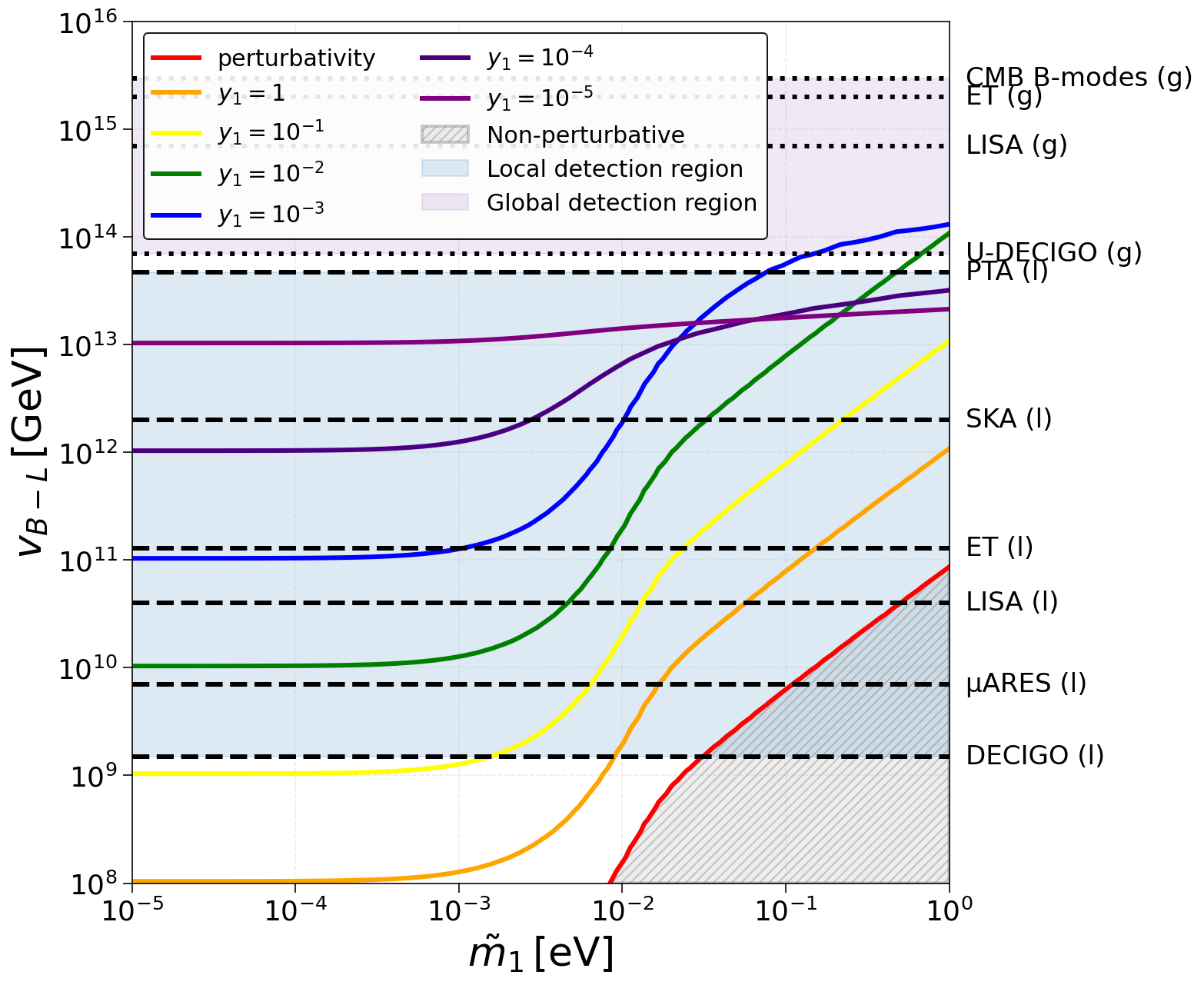}
\caption{\it Same as the plot above only for non-thermal initial abundance of $\phi$ for $Y_\phi^i=0.1$ (\textbf{Left}) and $Y_{\phi}^i=10^{-2}$ (\textbf{Right}). We can see the larger initial abundances lower the scale of leptogenesis in the weak washout regime however for the strong washout regime for large Yukawas we reproduce the thermal leptogenesis results. This is due to the strong washout regime being independent of initial conditions.}
    \label{fig:vblscanfull2}
\end{figure}
As expected the larger the Yukawa and the initial abundance the lower the scale of leptogenesis can be in the weak washout regime. However for larger $\tilde m_1$ and $y_1$ values the results reproduce that of the thermal initial conditions. This reflects the known result that in the strong washout regime the baryon asymmetry produced is independent of initial conditions \cite{Giudice:2003jh}.\\
Local strings probe most of the testable region of parameter space for thermal leptogenesis in contrast to global strings which only probe the high mass region. We note that once one includes such scalar fields needed to facilitate a spontaneously broken B-L symmetry, it is not quite true that next generation gravitational wave probes of cosmic string networks are sensitive to the full range relevant to thermal leptogenesis. For both thermal and non-thermal leptogenesis in the strong washout regime however local strings can probe the entire parameter space of interest.

\medskip

\section{Discussion and Conclusions}
\label{sec: conclusion}
We have carried out a detailed study of high and low-scale leptogenesis within the type I seesaw framework in the presence of a $U(1)_{B-L}$ symmetry. While related scenarios have been considered previously, our study incorporates several elements that have not been touched upon before with significant implications. These specific additions are summarised in Table~\ref{tab:Testing_Gauged_Global_Seesaw_Leptogenesis}. We first reviewed how breaking a high-scale $B-L$ symmetry produces cosmic strings that, in turn, sourced a potentially observable gravitational-wave background, separately for global and gauged cases. \par We then derived analytic conditions  which enabled us to identify the parameter space involving $( M_1, \tilde m_1 )$, for successful baryogenesis in section \ref{sec:analytic leptogenesis}. The thermal leptogenesis analysis yielded a lower bound on the lightest sterile neutrino mass, $M_1 > 1.74\times10^{8}\,\text{GeV}$ which is roughly an order of magnitude below the standard vanilla bound. We also examined non-thermal initial conditions, for which the corresponding requirement became $M_1 > 1.25\times10^{6}/Y_\phi^i\,\text{GeV}$ substantially reducing the scale necessary for successful leptogenesis. These bounds were found to hold in the regime $\tilde m_1\lesssim 10^{-4.5}$ eV.\\
Flavour-dependent washout effects yielded a precise best-fit expression for the efficiency of leptogenesis in equation \ref{eq:bestfit_thermal} and the minimal sterile-neutrino mass compatible with successful leptogenesis. We showed that, unlike in vanilla leptogenesis, the mass bound remained relatively mild: for any effective neutrino mass, it increased only to $2.4\times 10^8$ GeV shown in figure \ref{fig: high scale scan}. This occurs due to the late-decaying scalar field $\phi$ (also responsible for $U(1)_{B-L}$ symmetry breaking) which produced most of the right-handed neutrinos after washout processes had frozen out enabling for a much higher efficiency never going below $\kappa=0.7$. This is several orders of magnitude lower than the corresponding vanilla bound in the same regime. We carried out an analogous analysis for several non-thermal initial abundances, and the resulting best-fit expressions are summarised in Table~\ref{tab:kappa_fits} and minimum Mass bounds shown in figure \ref{fig: high scale scan nontherm}. We then extended the analysis to the near-resonant regime, showing in figure \ref{fig: near resonant scan} that successful baryogenesis could occur down to the TeV scale for any effective neutrino mass $\tilde m_1$.\\
Finally, we identified the regions of parameter space consistent with successful leptogenesis for fixed $v_{B-L}$ and determined which of these regions would be detectable through the gravitational-wave signals of global or gauged cosmic strings. The results for thermal leptogenesis and the various detector sensitivities are shown in figure \ref{fig:vblscanfull} and for non-thermal abundances in Figure \ref{fig:vblscanfull2}. A future detection or absence of a gravitational-wave background will therefore serve as a decisive test of this region of parameter space. We found that including flavour effects rescued some regions of the parameter space that were believed to be ruled out already in the strong washout regime.\\
It is worth revisiting the correlation between cosmic strings and thermal leptogenesis. Since GUT scale right handed neutrinos are in tension with perturbative couplings, the hierarchy between the seesaw-leptogenesis scale and the GUT scale warrants and explanation which is provided if there is a symmetry related to B-L. The majority of symmetry breaking paths that places inflation in between the production of phenomenologically dangerous monopole production and leptogenesis result in the production of a network of cosmic strings. In previous work it was argued that the entire range of parameter space relevant to thermal leptogenesis left a detectable signal. We, unfortunately, note that in this work, even though we have thermal initial conditions and a reheating tempetature above the mass of the right handed neutrino, scalar decays lower the minimum scales involved such that it is not quite true that next generation gravitational wave probes of cosmic string networks are sensitive to the full range relevant to thermal leptogenesis. Probing this paradigm would require a modest improvement in the sensitivity of next generation detectors. This is of course neglecting the near-resonant and resonant cases. \par
\begin{table}[h!]
\centering
\renewcommand{\arraystretch}{1.4}
\setlength{\tabcolsep}{12pt}
\begin{tabular}{|c|c|c|}
\hline
\textbf{GW from CS Testing the scale of} & \textbf{Gauged and Global} $U(1)_{B-L}$\\
\hline
\textbf{Flavoured Thermal Leptogenesis} & Figure \ref{fig:vblscanfull}\\
\hline
\textbf{Flavoured Non-Thermal Leptogenesis} & Figure \ref{fig:vblscanfull2} \\
\hline
\textbf{Near-resonant Leptogenesis} & Figure \ref{fig: near resonant scan}\\
\hline
\end{tabular}
\caption{\it Summary of results as relating to Table \ref{tab:Testing_Gauged_Global_Seesaw_Leptogenesis} and how this fulfils holes in the current literature of using gravitational waves from cosmic strings to probe leptogenesis.}
\label{tab:Testing_Gauged_Global_Seesaw_Leptogenesis2}
\end{table}
As a future outlook, we note that despite the scales of physics involve we would like to note the potential complementarity with a myriad of probes of baryogenesis via leptogenesis scenarios. Specifically, tools like CMB spectral indices~\cite{Ghoshal:2022fud}, domain walls~\cite{Barman:2022yos, King:2023cgv}, nucleating and colliding vacuum bubbles during first-order~\cite{Dasgupta:2022isg,Borah:2022cdx}, graviton bremmstrahlung~\cite{Ghoshal:2022kqp}, gravitational production of neutrinos~\cite{Haque:2023zhb}, inflationary tensor perturbations propagating as GW~\cite{Berbig:2023yyy,Borboruah:2024eal,Chianese:2025mll, Chianese_2024}, primordial blackholes~\cite{Perez-Gonzalez:2020vnz,Datta:2020bht,Barman:2024slw,JyotiDas:2021shi,Barman:2021ost,Bernal:2022pue,Bhaumik:2022pil,Bhaumik:2022zdd,Ghoshal:2023fno}, and measurements of primordial non-gaussianity  \cite{Cui:2021iie,Fong:2023egk} could play a key role in testing the see-saw and leptogenesis paradigms with the full picture of how testable this scenario is still yet to be finalized.
\medskip

\section*{Acknowledgements}
GW and AS acknowledge the STFC Consolidated Grant ST/X000583/1. AS thanks the University of Southampton School of Physics and Astronomy for the support of a Mayflower PhD scholarship. A.G. acknowledges the support from the Royal Society, UK, Funding Reference: NIF\ R1\ 253963.

\bibliographystyle{unsrt}
\bibliography{main}
\appendix
\section{Gauge Interactions}
\label{sec:appendix}
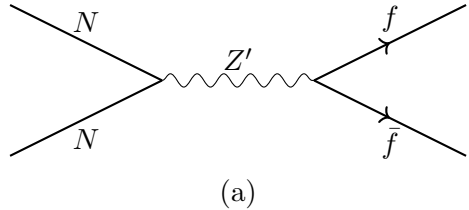
\begin{figure}[h!]
\centering
\begin{tikzpicture}[decoration={markings,mark=at position 0.5 with {\arrow{>}}}]
  \coordinate (N1) at (-3,1);
  \coordinate (N2) at (-3,-1);

  \coordinate (v1) at (-1,0);

  \coordinate (v2) at (1,0);

  \coordinate (f) at (3,1);
  \coordinate (fbar) at (3,-1);

  \draw[thick] (N1) -- (v1) node[midway, above] {$N$};
  \draw[thick] (N2) -- (v1) node[midway, below] {$N$};

  \draw[decorate,decoration=snake] (v1) -- (v2) node[midway, above] {$Z'$};

  \draw[thick,postaction={decorate}] (v2) -- (f) node[midway, above] {$f$};
  \draw[thick,postaction={decorate}] (v2) -- (fbar) node[midway, below] {$\bar f$};

  \node at (0,-1.5) {(a)};
\end{tikzpicture}
\caption{\it Annihilation of right-handed neutrinos into SM fermions via a $Z'$ gauge boson.}
\label{fig:NNToffbar}
\end{figure}

We now consider the gauge interactions and show that for our scales they are negligible. In the local theory, right-handed neutrinos interact through the $U(1)_{B-L}$ gauge boson, and can annihilate into Standard Model fermions. If this annihilation rate is comparable to the canonical decay of right-handed neutrinos into higgs and leptons this will hinder leptogenesis as fewer decays occur. We compute the annihilation rate $N N \to f\bar f$ mediated by a $Z'$ gauge boson of mass $M_{Z'} = 2 g v_{B-L}$, and show that for high-scale symmetry breaking ($v_{B-L}\gtrsim 10^8\,\mathrm{GeV}$) the process is always negligible. The process in question is shown in fig \ref{fig:NNToffbar}.

The annihilation proceeds through off--shell $Z'$ exchange and is described
by the reduced cross section
\begin{equation}
\label{eq:sigmahat_recall}
    \hat{\sigma}(s)
    = \frac{g_{B-L}^4\,s}{12\pi}
      \frac{\left(1-\frac{4M_N^2}{s}\right)^{3/2}}
      {\left(s-M_{Z'}^2\right)^2 + M_{Z'}^2\Gamma_{Z'}^2}
      \sum_f Q_{B-L}(f)^2 ,
\end{equation}
from which the thermally averaged cross section follows:
\begin{equation}
\label{eq:sigmav_thermal}
    \langle \sigma v \rangle (T)
    =
    \frac{T}{32\pi^4 \left[n_N^{\rm eq}(T)\right]^2}
    \int_{4M_N^2}^{\infty}
      ds\,\hat{\sigma}(s)\,\sqrt{s}\,
      K_1\!\left(\frac{\sqrt{s}}{T}\right).
\end{equation}
The corresponding annihilation rate per particle is
\begin{equation}
\label{eq:GammaAnn}
    \Gamma_{\rm ann}(T)
    = n_N(T)\,\langle \sigma v \rangle (T) ,
\end{equation}
where $n_N(T)$ is the actual neutrino abundance.
In the context of establishing when annihilations may affect leptogenesis. We evaluate this rate at $T=M_N$, and using a relativistic number density abundance. Once the right-handed neutrinos become non-relativistic this process will be bolztmann suppressed so if it is negligible while relativistic we will have shown it is negligible for leptogenesis.\\
A quantitative measure of the competition between the two processes is
given by the dimensionless ratio
\begin{equation}
\label{eq:R_def}
    R \equiv
    \frac{2 \Gamma_{\rm ann}(T=M_N)}
         {\Gamma_{\rm dec}(T=M_N)}.
\end{equation}
When $R\ll 1$ the $Z'$--mediated annihilation is irrelevant for the evolution of the $N$ abundance and can be safely neglected. The factor of two comes from there being two right-handed neutrinos annihilated in the process.
When $R\sim 1$ the two channels compete and the asymmetry produced by
$N\to \ell H$ decays is suppressed, while $R\gg 1$ corresponds to a
regime in which leptogenesis fails altogether.
In what follows we evaluate the ratio \eqref{eq:R_def} across the
$(g,y)$ parameter space and determine the region in which gauge
annihilations may be ignored. We scan over our ranges of $g,\ y\in [10^{-7}, 1]$ and take the smallest $v_{B-L}=10^6$ GeV that we are considering to maximise the annihilation process and we plot the ratio below in fig \ref{fig:gaugescan}, 
\begin{figure}[H]
    \centering
    \includegraphics[width=0.75\textwidth]{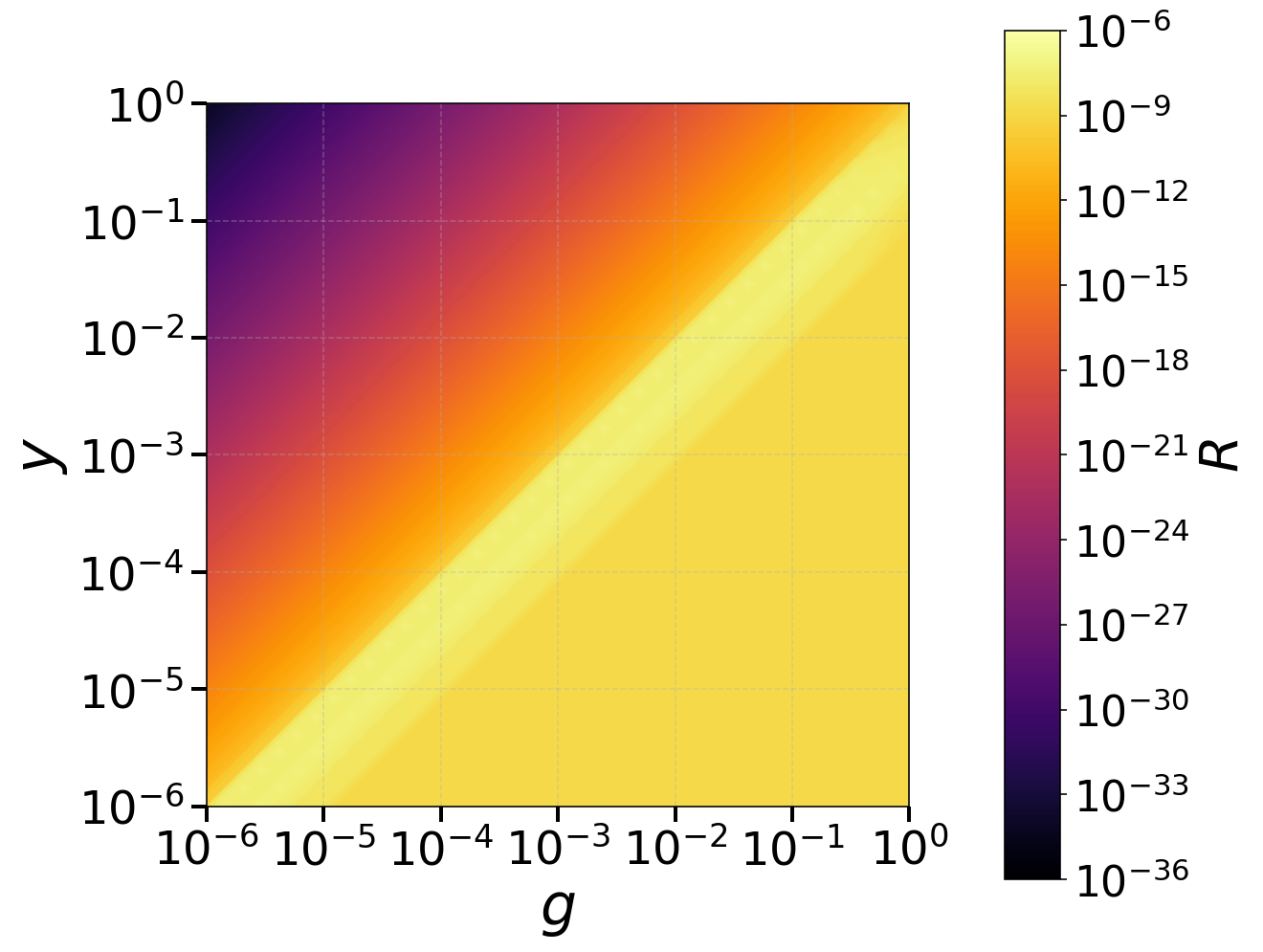}
    \caption{\it Scan of annihilation rate over decay rate taking the largest possible annihilation rate to decay rate ratio. we consider and the smallest decay rate we consider. As we can see the ratios are negligible meaning this process can safely be neglected}
    \label{fig:gaugescan}
\end{figure}
In all the parameter space we consider therefore these interaction can safely be neglected. For $v_{B-L}=10^7$ GeV these can also be safely neglected however for $v_{B-L}\lesssim 10^6$ GeV these effects can only be neglected for $y>g$ and kill leptogenesis if this condition is not satisfied as annihilations become dominant.\\
The other effect to consider is the production of right-handed neutrinos from $Z'$ decays if they can happen. As most of the decays will be into standard model fermions and the branching ratio into the lightest sterile neutrino is small however and we neglect this. 
\end{document}